\documentclass[journal]{IEEEtran} 

\ifCLASSINFOpdf
    \usepackage{graphicx,color}  
\else
\fi

\usepackage{amsmath,amssymb,amsfonts} 
\usepackage{cleveref} 
\Crefformat{figure}{#2Fig.~#1#3}
\Crefmultiformat{figure}{Figs.~#2#1#3}{ and~#2#1#3}{, #2#1#3}{ and~#2#1#3}
\usepackage[thmmarks,amsmath]{ntheorem}
\theoremnumbering{arabic}
\theoremstyle{break}
\theoremsymbol{}
\theoremheaderfont{\normalfont\bfseries}
\theorembodyfont{\upshape}
\newtheorem{remark}{Remark}
\theoremstyle{nonumberbreak}

\usepackage[
]{glossaries}
\newacronym{3gpp}{3GPP}{Third Generation Partnership Project}
\newacronym{2g}{2G}{Second Generation}
\newacronym{5g}{5G}{Fifth Generation}
\newacronym{acsi}{ACSI}{Absolute Channel State Information}
\newacronym{aoa}{AOA}{angle-of-arrival}
\newacronym{aod}{AOD}{angle-of-departure}
\newacronym{ann}{ANN}{Artificial Neural Network}
\newacronym{asic}{ASIC}{application specific integrated circuit}
\newacronym{awgn}{AWGN}{additive white Gaussian noise}
\newacronym{bdir}{BDIR}{Beam-Domain Impulse Response}
\newacronym{bs}{BS}{Base Station}
\newacronym{ccs}{CCS}{correlative channel sounder}
\newacronym{ccsi}{CCSI}{Complete Channel State Information}
\newacronym{ce}{CE}{channel estimation}
\newacronym{cnn}{CNN}{Convolutional Neural Network}
\newacronym{csi}{CSI}{Channel State Information}
\newacronym{csp}{CSP}{Contact Service Point}
\newacronym{crlb}{CRLB}{Cram\'er-Rao lower bound}
\newacronym{crs}{CRS}{Cell-specific Reference Signals}
\newacronym{cvnn}{CVNN}{Complex-Valued Neural Network}
\newacronym{da}{DA}{data association}
\newacronym{dc}{DC}{direct current}
\newacronym{dm}{DM}{diffuse multipath}
\newacronym{dmc}{DMC}{\gls{dm} component}
\newacronym{ecsp}{ECSP}{Edge Computing Service Point}
\newacronym{eirp}{EIRP}{equivalent isotropically radiated power}
\newacronym{en}{EN}{energy neutral}
\newacronym{end}{END}{energy neutral device}
\newacronym{epu}{EPU}{edge processing unit}
\newacronym{fa}{FA}{federation anchor}
\newacronym{fcnn}{FCNN}{Fully-Connected Neural Network}
\newacronym{fdd}{FDD}{Frequency Division Duplexing}
\newacronym{ft}{FT}{Fourier Transform}
\newacronym{gbsc}{GBSC}{geometry-based deterministic stochastic channel}
\newacronym{gdop}{GDOP}{Geometric Dilution of Precision}
\newacronym{gnss}{GNSS}{Global Navigational Satellite Systems}
\newacronym{gps}{GPS}{Global Positioning System}
\newacronym{gpst}{GPST}{GPS Time}
\newacronym{gsm}{GSM}{Global System for Mobile Communications}  %
\newacronym{heu}{HEU}{High Epistemic Uncertainty}
\newacronym{ic}{IC}{integrated circuit}
\newacronym{id}{ID}{information decoding}
\newacronym{iot}{IoT}{Internet of Things}
\newacronym{knn}{kNN}{k-Nearest Neighbor}
\newacronym{leu}{LEU}{Low Epistemic Uncertainty}
\newacronym{los}{LoS}{line-of-sight}
\newacronym{lsa}{LSA}{large synthetic array}
\newacronym{lte}{LTE}{Long Term Evolution}
\newacronym{lti}{LTI}{linear time-invariant}
\newacronym{mmwave}{mmWave}{millimeter wave}
\newacronym{meo}{MEO}{Medium Earth Orbit}
\newacronym{mfad}{MFAD}{Matched Filter Azimuthal-Delay}
\newacronym{mfpd}{MFPD}{Matched Filter Polametric-Delay}
\newacronym{mimo}{MIMO}{Multiple-Input Multiple-Output}
\newacronym{miso}{MISO}{multiple-input single-output}
\newacronym{mpc}{MPC}{Multipath Component}
\newacronym{nlp}{NLP}{Natural Language Processing}
\newacronym{ofdm}{OFDM}{orthogonal frequency-division multiplexing}
\newacronym{ota}{OTA}{over-the-air}
\newacronym{nb}{NB}{narrowband}
\newacronym{nlos}{NLoS}{non-line-of-sight}
\newacronym{pa}{PA}{power amplifier}
\newacronym{pl}{PL}{path loss}
\newacronym{pc}{PC}{pilot count}
\newacronym{pss}{PSS}{Primary Synchronization Signal}
\newacronym{rcs}{RCS}{radar cross section}
\newacronym{rf}{RF}{radio frequency}
\newacronym{rfid}{RFID}{radio frequency identification}
\newacronym{ris}{RIS}{reflective intelligent surface}
\newacronym{rms}{RMS}{root mean square}
\newacronym{rnn}{RNN}{Recurrent Neural Network}
\newacronym{rss}{RSS}{Received Signal Strength}
\newacronym{rssi}{RSSI}{received signal strength indicator}
\newacronym{ra}{RA}{reflectarray}
\newacronym{rtk}{RTK}{Real-Time Kinematic}
\newacronym{rw}{RW}{RadioWeaves}
\newacronym{sa}{SA}{synchronization anchor}
\newacronym{sar}{SAR}{specific absorption rate}
\newacronym{sbl}{SBL}{sparse Bayesian learning}
\newacronym{sdn}{SDN}{software-defined network}
\newacronym{sinr}{SINR}{signal-to-interference-plus-noise ratio}
\newacronym{simo}{SIMO}{single-input multiple-output}
\newacronym{siso}{SISO}{single-input single-output}
\newacronym{slam}{SLAM}{simultaneous localization and mapping}
\newacronym{snr}{SNR}{signal-to-noise ratio}
\newacronym{smc}{SMC}{specular multipath component}
\newacronym{sss}{SSS}{Secondary Synchronization Signal}
\newacronym{suca}{SUCA}{Stacked Uniform Circular Array}
\newacronym{s-parameters}{S-parameters}{scattering parameters}
\newacronym{tosm}{TOSM}{through-open-short-match}
\newacronym{tdoa}{TDOA}{time-difference-of-arrival}
\newacronym{toa}{TOA}{time-of-arrival}
\newacronym{ue}{UE}{User Equipment}
\newacronym{uhf}{UHF}{ultra high frequency}
\newacronym{ula}{ULA}{uniform linear array}
\newacronym{upa}{UPA}{uniform planar array}
\newacronym{ura}{URA}{uniform rectangular array}
\newacronym{uwb}{UWB}{ultrawideband}
\newacronym{vhf}{VHF}{Very High Frequency}
\newacronym{vlf}{VLF}{Very Low Frequency}
\newacronym{vna}{VNA}{vector network analyzer}
\newacronym{wb}{WB}{wideband}
\newacronym{wpt}{WPT}{wireless power transfer}
\newacronym{zf}{ZF}{zero forcing}

\newacronym{cdf}{CDF}{cumulative distribution function}
\newacronym{ssf}{SSF}{small-scale fading}
\newacronym{lsf}{LSF}{large-scale fading}
\newacronym{ple}{PLE}{pathloss exponent}
\newacronym{usrp}{USRP}{universal software radio peripheral}
\newacronym{rb}{Rb}{Rubidium}
\newacronym{pps}{1PPS}{1 pulse-per-second}
\newacronym{mrc}{MRC}{maximum ratio combinig}
\newacronym{fpga}{FPGA}{field-programmable gate array}
\newacronym{agc}{AGC}{automatic gain control}
\newacronym{tdma}{TDMA}{time-division multiple access}
\newacronym{cp}{CP}{cyclic prefix}
\newacronym{dac}{DAC}{digital-to-analog converter}

\newacronym{pdf}{PDF}{probability density function}
\newacronym{cbsm}{CBSM}{correlation-based stochastic model}
\newacronym{gbsm}{GBSM}{geometry-based stochastic model}
\newacronym{vr}{VR}{visibility region}
\newacronym{acf}{ACF}{autocorrelation function}

\newacronym{jcas}{JCAS}{joint communication and sensing}
\newacronym{v2x}{V2x}{vehicle-to-everything}
\newacronym{m2m}{M2M}{machine-type communication}
\newacronym{dmimo}{D-MIMO}{distributed MIMO}
\newacronym{urllc}{URLLC}{ultra-reliable and low-latency communication}

\def\BibTeX{{\rm B\kern-.05em{\sc i\kern-.025em b}\kern-.08em
    T\kern-.1667em\lower.7ex\hbox{E}\kern-.125emX}}
\AtBeginDocument{\definecolor{tmlcncolor}{cmyk}{0.93,0.59,0.15,0.02}\definecolor{NavyBlue}{RGB}{0,86,125}}

\begin{document}
\bstctlcite{IEEEexample:BSTcontrol}

\markboth{Wiometrics: Comparative Performance of Artificial Neural Networks for Wireless  Navigation}{Whiton {et al.}}

\title{Wiometrics: Comparative Performance\\of Artificial Neural Networks for\\ Wireless  Navigation}

\author{Russ Whiton,~\IEEEmembership{Student Member,~IEEE}, Junshi Chen,~\IEEEmembership{Student Member,~IEEE}, Fredrik~Tufvesson,~\IEEEmembership{Fellow,~IEEE}

\thanks{Manuscript received August 31, 2023; revised YYYYY ZZ, 2023.}
\thanks{Russ Whiton (Corresponding Author) is affiliated with Lund University and Volvo Car Corporation. email: (russell.whiton@volvocars.com, russell.whiton@eit.lth.se). Junshi Chen is affiliated with Lund University and Terranet. email: (junshi.chen@terranet.se, junshi.chen@eit.lth.se). Fredrik Tufvesson is affiliated with Lund University. email: (fredrik.tufvesson@eit.lth.se)}
\thanks{This work was supported by the Swedish Innovation Agency VINNOVA through the MIMO-PAD Project (Reference number 2018-05000). Computational resources were provided by the National Academic Infrastructure for Supercomputing in Sweden (NAISS) at Tetralith, partially funded by the Swedish Research Council through grant agreement no. 2022-06725.}
\thanks{Digital Object Identifier XX.XXXX/XXX.XXXX.XXXXXXX}}
\maketitle


\begin{abstract}
Radio signals are used broadly as navigation aids, and current and future terrestrial wireless communication systems have properties that make their dual-use for this purpose attractive. Sub-6 GHz carrier frequencies enable widespread coverage for data communication and navigation, but typically offer smaller bandwidths and limited resolution for precise estimation of geometries, particularly in environments where propagation channels are diffuse in time and/or space. Non-parametric methods have been employed with some success for such scenarios both commercially and in literature, but often with an emphasis on low-cost hardware and simple models of propagation, or with simulations that do not fully capture hardware impairments and complex propagation mechanisms. In this article, we make opportunistic observations of downlink signals transmitted by commercial cellular networks by using a software-defined radio and massive antenna array mounted on a passenger vehicle in an urban non line-of-sight scenario, together with a ground truth reference for vehicle pose. With these observations as inputs, we employ artificial neural networks to generate estimates of vehicle location and heading for various artificial neural network architectures and different representations of the input observation data, which we call wiometrics, and compare the performance for navigation. Position accuracy on the order of a few meters, and heading accuracy of a few degrees, are achieved for the best-performing combinations of networks and wiometrics. Based on the results of the experiments we draw conclusions regarding possible future directions for wireless navigation using statistical methods.  
\end{abstract}


\maketitle

\IEEEpeerreviewmaketitle

\section{Introduction}\label{sec_intro}


\IEEEPARstart{W}{ireless} systems are used ubiquitously for navigation. \gls{gnss} including the US Department of Defense's \gls{gps} deploy satellites in \gls{meo} which allow users anywhere on earth with an unobstructed view of the sky to generate estimates of global position and time with meter-level and microsecond-level accuracies using inexpensive mass-market hardware \cite{teunissen2017springer}. \gls{rtk} methods have been developed to enhance \gls{gnss} accuracy to just a few centimeters \cite{janssen2009comparison}, and millimeter-level accuracy is even achievable for scientific applications including measurement of tectonic plate velocities with sophisticated hardware and post-processing algorithms \cite{larson1997global}. 


With this impressive performance in mind, one could consider \gls{meo} navigational satellites to have solved the problem of positioning on Earth. However, numerous applications of interest face practical limitations that preclude using \gls{gnss} as the sole source for navigation. \gls{gnss} signal reception is limited to non-existent indoors \cite{kjaergaard2010indoor}, for example, and in urban environments observations are frequently not of sufficient quality to meet application requirements for accuracy, availability, continuity, or integrity \cite{zhu2018gnss}. More conspicuously, \gls{gnss} receivers are vulnerable to intentional and unintentional jamming, as well as deliberate spoofing \cite{psiaki2016gnss}. Augmentation with complementary proprioceptive sensing such as odometry, inertial measurement, or altimeters can help mitigate some of these limitations (dead-reckoning), but even sensor-fused systems still have utility for absolute navigation estimates in a global frame (position fixing) \cite{GrovesPrinciples}. 

Terrestrial wireless systems are particularly well-suited to provide global position estimates \cite{whiton2022cellular}, and cellular communications systems boast a decades-long history of dual-use for navigation driven primarily by legislation intended to assist first responders in emergencies \cite{1g_to_5g_cellular_localization}. Such systems have a natural affinity for use in navigation owing to their ubiquity of deployment and the large bandwidths and link budgets that enable high data-rate communication \cite{kassas_magazine}. \gls{5g} systems from the \gls{3gpp} have dedicated measurements for performing triangulation and multi-lateration \cite{dwivedi2021positioning}, measurements which are expected to offer unprecedented accuracy compared with what previous cellular systems could achieve. 

However, the same physical obstructions which prevent utilization of satellite signals in some environments can cause similar challenges for navigation systems utilizing terrestrial transmitters. Position fixing methods based on ranging or bearing (angular) measurements will face difficulties in the face of multipath, manifesting as positive range biasing \cite{muller2016statistical} and angular biases \cite{rusti_ion_2022}. A broad set of methods have been proposed for \gls{los} identification and multipath mitigation, including multipath-estimating tracking loops \cite{liu2022multipath}, statistical methods \cite{orabi2021machine}, and residuals testing \cite{maaref2021autonomous}. An even more promising solution for multipath is to exploit the additional information that it provides to improve performance rather than trying to mitigate it \cite{ChSLAMOrig, 5gfoetofriend}. Tracking of individual multipath components allows for the construction of multiple ranging and bearing observations from a single transmitter, but this too is not without challenges. Such methods are best suited for large channel bandwidths, antenna arrays, dense transmitter infrastructure, and limited mobility, or dense multipath may preclude reliable ranging and bearing measurements \cite{jiang2022survey}. 


 Feature-matching \cite{GrovesPrinciples} (or \textit{non-parametric}) navigation techniques rely on comparisons of observations with databases rather than using explicit calculation of ranges and bearing to known landmarks. With wireless navigation, these methods are referred to as \textit{proximity} or \textit{fingerprinting}, and are often used for lower carrier frequencies \cite{sharp2019wireless, wireless_future}. Billions of devices are in use today that can provide users with a coarse location based on a non-parametric ``network-provided" location, using massive databases of Wi-Fi and cellular transmitters aggregated from many users over a long time \cite{nedelkov2020accuracy}. The breakthrough success of machine learning in areas like computer vision and natural language processing has inspired significant recent interest in applying statistical learning techniques, primarily \glspl{ann} to wireless positioning \cite{survey_ML_wireless}, in order to make the leap from the coarse position provided by commercial systems to highly precise estimates. Such techniques are frequently predicated  on the use of massive-\gls{mimo} \cite{marzetta2016fundamentals}, in which multiple transmitting and/or multiple receiving antennas provide separate observations for a single link.

In this work, a massive-\gls{mimo} receiver is used to provide navigation state estimates for a passenger vehicle in a complicated urban propagation environment, building on previous work from the authors in \cite{whiton2022urban}. \glspl{ann} are used to estimate navigation states based on opportunistic measurements from commercial \gls{lte} \glspl{bs}. Wireless fingerprinting\footnote{Somewhat confusingly, ``fingerprinting" is also used in literature for performing transmitter identification by capturing characteristic device-level hardware variations \cite{merchant2018deep}. In this work, fingerprinting refers to non-parametric feature matching for navigation.} is analyzed broadly, and we offer the humble suggestion that this method might more appropriately be renamed \textit{wiometric navigation}, analogous to \textit{biometrics}, for which human fingerprinting is merely one tool (see Section~\ref{sec_ch_foot} for more exposition). This manuscript includes a number of extensions beyond our previous work in \cite{whiton2022urban}, including the following:
\begin{itemize}
    \item Multiple channel representations and \gls{ann} architectures are proposed and evaluated, as well as two train/test splits representing high and low levels of epistemic uncertainty.
    \item The navigation state and learning models are updated to include estimates of heading, as opposed to \cite{whiton2022urban}, which used heading as a network input rather than a predicted state at the output.
    \item Performance of the network is analyzed when signals are received concurrently from a neighboring \gls{bs}, and an architecture is proposed for additional sectors and \glspl{bs}.
\end{itemize}

The manuscript is organized as follows:  
Section~\ref{sec_ch_foot} discusses channel fingerprinting and specifies the four channel representations (wiometrics) used in the main body of the paper;  Section~\ref{sec_larnin_fw} discusses the types of algorithms used for association between the surveyed data and inferences at run-time; Section~\ref{sec_veh_meas} discusses the \gls{lte} signal structure, the vehicular test bed, and the data set with the two training and test splits; Section~\ref{sec_results} provides the results for the channel measurements, train/test splits and the neural networks; Section~\ref{sec_discussion} offers exposition on the results; and finally the conclusions are drawn in Section~\ref{sec_conclusion} together with suggestions for future work.

Notes on mathematical representation:
\begin{itemize}
    \item $(\cdot)^T$, $(\cdot)^H$, and $(\cdot)^*$ represent the transpose, Hermitian transpose and conjugate of a complex-valued vector/matrix, respectively.
    \item $\mathcal{F}_{m,n}$ indicates a two-dimensional $m$- and $n$-point \gls{ft}.
    \item $\mathbf{col_j(F)}$ and $\mathbf{row_k(F)}$ indicate taking the $i$-th column and the $k$-th row of a matrix, respectively.
\end{itemize}

\section{Channel Wiometrics}\label{sec_ch_foot}
\subsection{Channel and User State Representation}\label{ch_user_rep}

As formulated in \cite{joaovieira2017deep}, the intention with non-parametric wireless channel-based navigation methods is to learn an inverse function that maps a set of channel measurements $\mathbf{Y}_i$ to their associated navigation state vectors $\mathbf{x}_i$ in a maximally-bijective fashion:
\begin{equation}\label{fingerprinting_general}
    f^{-1} : \{\mathbf{Y}_i\} \rightarrow \{\mathbf{x}_i\}.
\end{equation}

Literature addressing this basic problem formulation can be found as early as 1993 \cite{prison_duress}, when a ``calibration matrix" was proposed to construct an effective radio map that compensated for the limitations that complex propagation phenomena imposed on angle-of-arrival methods in a prison environment for \gls{vhf} signals. Literature on the subject expanded to encompass other technologies such as WaveLAN \cite{bahl2000radar} and Wi-Fi, the well-known successor to WaveLAN \cite{khalajmehrabadi2017modern}. Application using cellular-based technology has been investigated from \gls{2g} systems \cite{gsm_fp} through \gls{5g} systems \cite{max_vtc}.

Much of this body of work on the subject has come from the Internet-of-Things domain, employing commercial hardware that is easy to use but which provides limited insight into the internal hardware states. \gls{rss} values were a logical starting point \cite{bahl2000radar}, reducing the dimensionality of the channel to a single scalar value corresponding to received power (a function of channel gain). A simple extension of this, which was enabled by network cards such as the Intel IWL-5300, was to consider the frequency-dependent channel gain or equivalently \gls{csi} \cite{deepfi}. \gls{csi}-based solutions to the problem formulation of \Cref{fingerprinting_general} are so thoroughly investigated that they have spawned dedicated survey papers \cite{liu2019survey_csi} and derivative methods including feature engineering of the \gls{csi} vectors \cite{pecoraro2018csi}, multi-antenna \gls{csi} \cite{chapre2015csi, arnold2019novel}, or stacking of subsequent CSI samples in the time domain \cite{chen2017confi}. We formulate the expression of \gls{csi} at each time index $i$ for $M$ receive antenna ports ($M = P * M_I * M_{II}$ for a two-dimensional antenna array with $P$ polarizations, $M_I$ first and $M_{II}$ second dimensions of the array) and $S$ frequency samples, typically \gls{ofdm} subcarriers\footnote{The measurement system described in Section~\ref{sec_veh_meas} utilizes \gls{lte} signals. We formulate our definitions accordingly as generic for \gls{ofdm} systems and note that use of \gls{ofdm} is likely to continue for many use cases even with the advent of 6G systems \cite{harsh_6g_paper}.}, as the complex-valued matrix $\mathbf{F}_i$:

\begin{equation}\label{eq:csi_basis}
\mathbf{F}_i  \in \mathbb{C}^{S\times M}.
\end{equation}

Numerous other representations have been tested that employ additional signal processing to produce representations that are intuitive or hypothesized to provide compelling performance gains in the face of aleatoric or epistemic uncertainty. The domains tested include angles of arrival \cite{wang2018deep}; channel impulse responses \cite{niitsoo2019deep, max_vtc}; delay spreads or number of multipath components \cite{ye2017neural}; hybrids of various measurements defined in 3GPP standards \cite{li2019deep_rsrpdude}; and angular-delay domain \cite{joaovieira2017deep, sun2018single, hejazi2021dyloc, wu2021learning}. 



The user navigation state representation $\mathbf{x}_i$ is similarly flexible. Most works on the subject consider a two-dimensional Cartesian position and formulate a two-dimensional regression problem, but it is feasible to parameterize user position discretely in a grid, by building and floor \cite{floor_determination, ferrand2020dnn}, or as any other kind of classification problem. Information about uncertainties (practical for many navigation problems) is also directly estimable via a network, as has been demonstrated through direct estimation \cite{li2019wireless} and through probability maps \cite{gonultacs2021csi}. The user states estimated in this paper are position and orientation in a two-dimensional plane: an East-West coordinate $x_e$, a North-South coordinate $x_n$, and a heading (or equivalently, yaw) value $x_{\gamma}$ in degrees ranging from $[-180^\circ , 180^\circ)$:

\begin{equation}\label{eq:cnn_eq}
\mathbf{x}_{i} = {[x_e, x_n, x_{\gamma}]}^T \in \mathbb{R}^3 .
\end{equation}

\begin{remark}
Some ``fingerprinting" literature for multi-antenna systems formulates a tracking problem, where inferences are made about a mobile agent's state based on multi-antenna measurements on the network side. The measurement system in this work (described in \Cref{sec_veh_meas}) employs a massive antenna array for a mobile receiver, which provides high-resolution observation of signals transmitted by a single antenna port from a \gls{bs}. In our previous work, we noted that user orientation is either interesting to use as an input to improve performance \cite{whiton2022urban} or to add as an estimable parameter if the channel representation is rich enough to allow for it \cite{rusti_ion_2022}.
\end{remark}

\subsection{Fingerprinting to Wiometrics}

In fingerprinting in the literal sense with human fingers, the goal is to effectively capture friction ridges on the fingers as a proxy for identity for the purpose of \textit{verification}, confirming a one-to-one comparison with a database entry, or for \textit{classification}, conducting a one-to-many match \cite{maltoni2009handbook}. The recorded pattern from the individual's fingers is analogous to $\mathbf{Y}_i$ in \Cref{fingerprinting_general}, and identity to $\mathbf{x}_i$. Channel fingerprinting can be formulated similarly as a \textit{proximity} problem with a binary one-to-one hypothesis, as a one-to-many database \textit{region classification} problem, or as a continuously-defined output space in multiple dimensions \cite{survey_ML_wireless}.

The analogy between channel fingerprinting and human fingerprinting is reasonably apt if judged by the performance criteria in each case. The proxy values $\mathbf{Y}_i$ should be as invariant as possible to environmental or measurement aperture changes; variability for wireless fingerprints is introduced by changes in scatterer geometry or differences in receiver hardware. For human fingerprint recording, variability can stem from non-uniform contact, variable skin condition, and dirty sensor plates \cite{tabassi2004fingerprint}. The proxy values in either case should be \textit{collectable}, in that the measurement apparatus should be reasonable in terms of hardware and software cost and complexity. Critically, the inherent challenge in developing reliable parametric models for human fingerprints or wireless channels in complex propagation environments means that statistical methods are attractive.

The analogy fails in that human fingerprints are a specific and unchanging trait for an individual, whereas wireless channel ``fingerprints" can be almost arbitrarily imaginative, as the numerous representations cited in the previous subsection demonstrate. The analogy to human fingerprinting may have made sense when hardware limitations meant that a scalar value for signal strength was the only easily realizable form for $\mathbf{Y}_i$ to take, but a more appropriate analogy is to \textit{biometric authentication}, of which human fingerprinting is just one subcategory \cite{springer_intro_biometrics}. Facial and voice recognition are also well-known biometric authentication methods, but even less well-known methods such as gait recognition or hand geometry recognition are also proxies for identity, useful when discretion or non-intrusiveness are desirable for inference. With this in mind, we suggest that all such non-parametric wireless navigation approaches (which do not explicitly calculate bearings and ranges to landmarks) should be renamed \textit{wiometric navigation}, and the underlying channel representations to \textit{wiometrics}, for wireless observation metric or simply wireless metrics in homage to biometrics. Four wiometrics are presented for analysis in the following subsection and examined for the duration of the manuscript.




\subsection{A Selection of Wiometrics for Analysis}

\begin{figure}[ht]
	\centering
	\includegraphics[width=8.5cm]{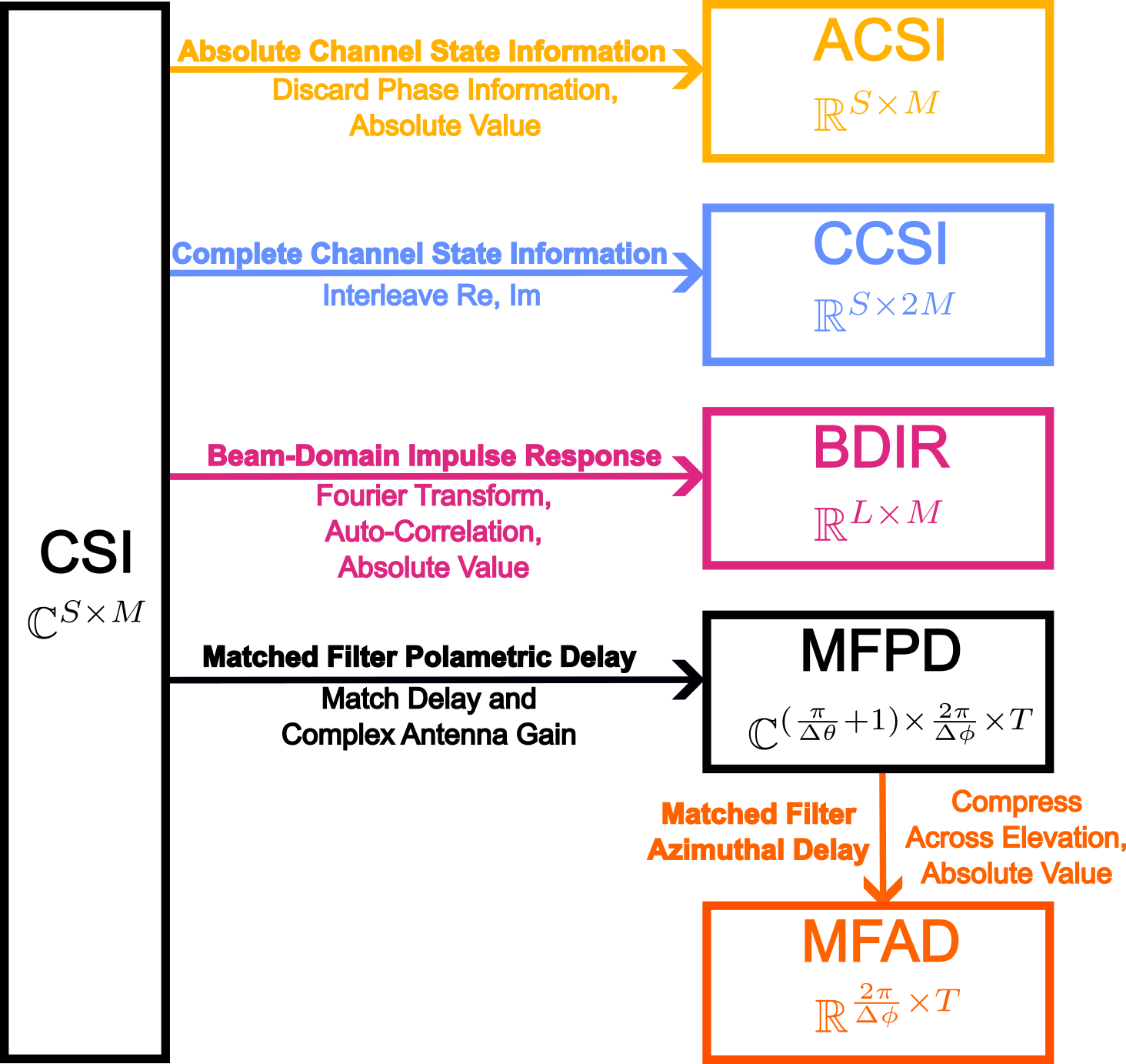}
	\caption{Block diagram of channel representations (wiometrics). Note that Channel State Information, the complex-valued transfer function for each antenna, is the basis for all subsequent representations.}
	\label{fig:chrepfig}
\end{figure}

Four channel representations are chosen for analysis, using the same measurement aperture with varying degrees of feature engineering, listed in ascending order of computational complexity and illustrated briefly in \Cref{fig:chrepfig}. As the figure shows, the \gls{csi} representation (complex-valued transfer function per antenna) is the basis for further signal processing to create real-valued representations\footnote{While \glspl{cvnn} have shown promise in the signal processing domain, they can be classified as an emerging field \cite{bassey2021survey} and are not considered in this work.} used for training \glspl{ann}.

For the remainder of the manuscript the subscript indicating time index $i$ is dropped for brevity, and elements of the matrix $\mathbf{F}$ are indicated with subscripts for the $j$-th row and $k$-th column as $f_{i,k}$.

\subsubsection{ACSI/CCSI - Absolute/Complete Channel State Information}\label{eq:xcsi_section}

As discussed in \Cref{ch_user_rep}, the frequency-dependent transfer function has long been used for non-parametric positioning methods, including the frequency response of multiple antennas for multi-receive antenna systems. The relationship of the phases among subcarriers is influenced by the complex antenna gain pattern as well as the total path length (delay) of individual multipath components. Taking the absolute value of the frequency response is equivalent to discarding that phase information, but this is frequently done for convenience (owing to the difficulty of implementing \glspl{cvnn}). The \gls{acsi} is therefore taken as a baseline for subsequent representations to be compared with, and denoted as $\mathbf{Y}_{ACSI}$:

\begin{equation}\label{y_acsi}
\mathbf{Y}_{ACSI} = | \ \mathbf{F} | \\\in \mathbb{R}^{S\times M}.
\end{equation}

Intuitively, $\mathbf{Y}_{ACSI}$ should not perform as well as systems that retain phase information because the absolute and relative phases of signals are the fundamental tool for calculation of ranges and bearings; \gls{gnss} uses code and carrier phase to set up the multi-lateration problem, for example. However, information about position and orientation can still be inferred through signal magnitude and the relationships among magnitudes in frequency and among antennas. 

An alternative representation to retain the phase information is to split out the real and imaginary components of $\mathbf{F}$ by interleaving on a column-by-column basis to form a new matrix, with twice the number of columns. Phase and amplitude are both maintained and no information is discarded, so the representation is referred to as \gls{ccsi} and $\mathbf{Y}_{CCSI}$ is defined as follows:

\begin{equation}\label{y_ccsi}
\resizebox{.9\hsize}{!}{$\mathbf{Y}_{CCSI} = \begin{bmatrix} 
Re(f_{1,1}) &  Im(f_{1,1}) & \dots  & Im(f_{1,M}) \\ 
Re(f_{2,1}) &  Im(f_{2,1}) & \dots  & Im(f_{2,M}) \\
\vdots      & \vdots       & \ddots & \vdots       \\
Re(f_{S,1}) &  Im(f_{S,1}) & \dots  & Im(f_{S,M}) \\ 
\end{bmatrix} \\\in \mathbb{R}^{S\times 2M}$}.
\end{equation}

The components are interleaved rather than stacking all real components and all imaginary components into sub-matrices to capture the relationships in phase for adjacent subcarriers with convolutional methods, discussed in \Cref{cnnbeskriv}. 


\subsubsection{BDIR - Beam-Domain Impulse Response}\label{eq:bdir_section}

Several methods have been proposed which exploit auto-correlation of received signals, under the hypothesis that auto-correlation across time, frequency, and/or space should serve as an effective proxy for position. If used as the input for an \gls{ann}, this can be regarded as a form of feature engineering. In \cite{studer2018channel}, the authors propose using a vectorized form of the matrix $\mathbf{F}$, by stacking the columns of the \gls{csi} matrix $\mathbf{F}$ into a ``radio geometry" vector $\mathbf{f}_{RG}$:

\begin{equation}
    \mathbf{f}_{RG} = \begin{bmatrix} 
    \mathbf{col_1(F)} \\ \mathbf{col_2(F)} \\
    \vdots \\ \mathbf{col_S(F)} 
    \end{bmatrix} \in \mathbb{C}^{(M*S)\times 1}.
\end{equation}

Taking the outer product $\mathbf{f}_{RG} * \mathbf{f}_{RG}^H$ ``lifts" the vectorized form to a higher dimensional feature geometry space which has dimensionality corresponding to the number of subcarriers and antennas multiplied $\mathbb{C}^{(M*S)\times (M*S)}$. This may subsequently be scaled or other transforms applied before dimensionality reduction (the Channel Charting function) is used to create a low-dimensional representation that is useful for network functions as an effective proxy for user position. This has been extended to be mixed with labeled data for enhanced positioning \cite{semisup_cc, Stahlke_tmlcn}. The enormous dimensionality of the ``space lifted" channel provides a large input space for pattern finding, but also entails computationally intensive matrix operations that grow as the square of the number of subcarriers and number of antennas. 

In \cite{ferrand2020dnn}, the authors also seek to do feature design for multi-antenna \gls{csi} in order to construct a channel representation that is robust against hardware impairments, computationally efficient, and which retains the essential information relevant for positioning. First, a two-dimensional \gls{ft} is taken across the two physical dimensions of a planar antenna array $M_I$ and $M_{II}$ for transformation into the beam domain. As with Channel Charting, auto-correlation is used, but a decimated frequency-domain version rather than across a space-lifted version, creating a shortened impulse response. The subsequent representation is referred to here as the \gls{bdir}.

More precisely, the beam-domain transformation entails first rearranging $\mathbf{F}$ into a four-dimensional array. The four-dimensional beam-domain version $\mathbf{Y}_ {BD,4D}$ is then formed by taking the \gls{ft} across the $M_I$ and $M_{II}$ dimensions of the two-dimensional antenna array\footnote{For the measurement platform described in \Cref{sec_veh_meas}, the two dimensions of the antenna array are considered as rings (four) and patches per ring (sixteen).}, and then rearranging back into the original shape along the same dimensions:


\begin{equation}\label{eq:bdir_ft}
\begin{aligned}
   \mathbf{F}  \in \mathbb{C}^{S\times M} & \rightarrow \mathbf{F}_{4D}  \in \mathbb{C}^{S \times M_I \times M_{II} \times P} \\
   \mathbf{Y}_ {BD,4D} & = \mathcal{F}_{M_I,M_{II}} \{ \mathbf{F}_{4D} \} \\
   \mathbf{Y}_ {BD,4D} & \rightarrow \mathbf{Y}_ {BD}  \in \mathbb{C}^{S \times M}.
\end{aligned}
\end{equation}



Auto-correlation is subsequently applied along the frequency domain of $\mathbf{Y}_ {BD}$ to remove the impact of timing advance and certain other hardware impairments. Both a decimation (downsampling) rate $\delta_{dec}$ (not every possible subcarrier spacing is considered) and a maximum number of adjacent frequency bins $\delta_{max}$ are used. In the original paper \cite{ferrand2020dnn}, a maximum subcarrier spacing of $\delta_{max} =64$ is used with a decimation rate of $ \delta_{dec} = 4$, resulting in $L = \lfloor\frac{\delta_{max}}{\delta_{rate}}\rfloor = 16$ subcarrier spacings $\delta_l \in [4, 8, ..., 64]$. The auto-correlation function $R_{\delta_l}$ for an offset of $\delta_l$ is expressed $R_{\delta_l}(\mathbf{y}) = \sum_{n=1}^{\delta_{max}} y(n)y^*(n+\delta_l)$ and applied for each antenna and time offset of $\mathbf{Y}_ {BD}$:



\begin{equation}\label{eq:bdir_matrix}
\resizebox{.91\hsize}{!}{$
{Y}_ {BDIR}  = \begin{bmatrix} 
    |R_{\delta_1} (\mathbf{col_1}(\mathbf{Y}_{BD}))| & \dots  & |R_{\delta_1} (\mathbf{col_M}(\mathbf{Y}_{BD}))| \\
    \vdots & \ddots & \vdots\\
    |R_{\delta_L} (\mathbf{col_1}(\mathbf{Y}_{BD}))| & \dots  & |R_{\delta_L} (\mathbf{col_M}(\mathbf{Y}_{BD}))|
    \end{bmatrix} \in \mathbb{C}^{L\times M}
$}.
\end{equation}

Note that the \gls{ft} across elements to formulate the beamspace representation in \Cref{eq:bdir_ft} does not use the measured antenna pattern. This is advantageous in terms of computational efficiency, and the invariance to certain hardware calibration errors and timing mismatch is useful for practical systems. However, for the sake of maximizing performance, one more wiometric is suggested in the following subsection.





\subsubsection{MFPD/MFAD - Matched Filter Polametric/Azimuthal-Delay}\label{eq:mfxd_section}

Decomposition of the channel into the temporal and spatial domains is appealing because it lends itself well to human intuition about what is happening as the geometry of the scenario changes. \gls{mpc} path lengths and angles-of-arrival/departure change intuitively with the geometry of the scattering environment. First proposed by \cite{joaovieira2017deep} for a two-dimensional simulated scenario, variations of this Angle-Delay Channel Amplitude/Power Matrix (ADCAM or ADCPM) \cite{sun2018single,sun2019fingerprint, wu2021learning, fast_singlesite} or Angle-Delay Profile (ADP) \cite{hejazi2021dyloc, chu2022sa}, have been popular in literature. 

We use the abbreviation \gls{mfpd} to emphasize that the complete three-dimensional radiation pattern per antenna port is used for formulating the angular response as in \cite{rusti_ion_2022}. The antenna response matrix for all antennas in discrete steps of $\Delta \theta$ for the elevation domain and azimuthal angle steps of $\Delta \phi$ is given as $\mathbf{A}$:

\begin{equation}\label{eq:ant_rad_pattern}
\mathbf{A}  \in \mathbb{C}^{ M\times ((\frac{\pi}{\Delta \theta} + 1)*(\frac{2\pi}{\Delta \phi}))}.
\end{equation}

A tacit assumption in the \gls{mfpd}/\gls{mfad} representation of the channel is that transmitter-receiver time synchronization is guaranteed and does not drift. In \cite{ferrand2020dnn}, the authors make the point that such a representation presumes absolute time synchronization between sender and receiver, with little allowance for timing advance, hardware impairments, or oscillator drift. Timing mismatch is discussed further in \Cref{sec_veh_meas}; for now we formulate the matched filter time-domain response per antenna and discrete time step $\tau$ for a maximum number of time steps $T$ as the matrix $\mathbf{D}$:

\begin{equation}\label{eq:time_matrix}
\mathbf{D}  = \begin{bmatrix} 
    1 &  e^{2 \pi f_1 \tau} & \dots  & e^{2 \pi f_1 \tau (T-1)}\\
    \vdots & \vdots & \ddots & \vdots\\
    1 & e^{2 \pi f_S \tau} & \dots  & e^{2 \pi f_S \tau (T-1)} 
    \end{bmatrix} \in \mathbb{C}^{S\times T}.
\end{equation}

Combining both the antenna and phase-delay elements of the matched filter response with the per-antenna and per-subcarrier matrix $\mathbf{F}$ yields the two-dimensional \gls{mfpd} matrix $\mathbf{Y}_{MFPD,2D}$:

\begin{equation}\label{eq:matched_filt}
\mathbf{Y}_{MFPD,2D} = \mathbf{A}^H \mathbf{F} \mathbf{D}^* \in  \mathbb{C}^{((\frac{\pi}{\Delta \theta} + 1)*(\frac{2\pi}{\Delta \phi})) \times T},
\end{equation}

\noindent which can be rearranged into a three-dimensional array  in the order of elevation, azimuth, and delay:

\begin{multline}\label{eq:threedee_mfad}
\mathbf{Y}_{MFPD,2D} \in  \mathbb{C}^{((\frac{\pi}{\Delta \theta} + 1)*(\frac{2\pi}{\Delta \phi})) \times T} \rightarrow \\ \mathbf{Y}_{MFPD} \in  \mathbb{C}^{(\frac{\pi}{\Delta \theta} + 1) \times (\frac{2\pi}{\Delta \phi}) \times T}.
\end{multline}


Summation (incoherent) across the elevation dimension (the $n$ rows in the first dimension $a$ of the three dimensions $abc$  of the absolute value of the $\mathbf{Y}_{MFPD}$ matrix) yields the \gls{mfad} representation, which is the wiometric used in the remainder of the manuscript. For a system with limited resolution in elevation (as many terrestrial systems have), the simplification of working in two dimensions with a smaller input is a desirable trade-off.  

\begin{equation}
    \mathbf{Y}_{MFAD} = \sum_{a=1}^{\frac{\pi}{\Delta \theta} + 1}  |\mathbf{Y}_{MFPD,abc}| \in \mathbb{R}^{(\frac{2\pi}{\Delta \phi}) \times T}.
\end{equation}


%
%
%








\section{Learning Frameworks}\label{sec_larnin_fw}
\subsection{Standard Methods and Deep Learning}

Feature-matching for navigation encompasses a broad family of technologies and methods, from Terrain-Referenced Navigation in aviation to gravity gradiometry on submarines \cite{GrovesPrinciples}. A recent comprehensive survey on machine learning with a particular focus on wireless navigation follows a common grouping convention that includes supervised, semi-supervised, and unsupervised methods \cite{survey_ML_wireless}. Within the supervised category two subcategories are identified. The first subcategory is ``Standard Methods" including \gls{knn}, Kernel Based Methods, Gaussian Process-based methods, and Trees/Ensemble methods. The second category is ``Deep Learning" methods, with particular emphasis on Deep Learning, which has been the primary solution for most learning problems in recent years\footnote{Deep Learning is technically a subcategory of \glspl{ann}, but few \glspl{ann} are employed that do not meet the criterion of being ``deep" (having hidden layers).}. We compared one ``standard" method (\gls{knn}) and deep learning in our previous work \cite{whiton2022urban}.  

Popular classifiers for computer vision competitions such as AlexNet \cite{krizhevsky2017imagenet} or ResNet \cite{he2016deep} and object classifiers such as YOLO \cite{Redmon_2017_CVPR} have inspired use of \glspl{ann} in virtually every domain (see, e.g., ``Applications of Machine Learning" in \cite{sarker2021machine}). An even more recent trend dominating machine learning literature, not mentioned in \cite{survey_ML_wireless}, is the trend of using Transformers \cite{vaswani2017attention}. This has dominated \gls{nlp} research, and has already attracted interest in computer vision \cite{khan2022transformers}. No Transformers are implemented in this work, but suggestions about their future potential for navigation are considered in the Discussion (\Cref{sec_discussion}).

\subsection{Artificial Neural Networks}\label{sec_fcnns}

\glspl{ann} come in many shapes, sizes, and architectures. Networks in computer vision have evolved toward larger and deeper networks for achieving breakthrough performance gains \cite{Redmon_2017_CVPR}. For wireless navigation using \glspl{ann}, hierarchical structures have been proposed to reduce total processing time, even for a single transmitter scenario \cite{fast_singlesite}. This section is not intended to be a comprehensive survey of recent \gls{ann} use for wireless navigation, but aims to provide a sampling of recent results; and, for a set of \glspl{ann} described in the following subsections, together with the wiometrics of \Cref{sec_ch_foot} and dataset generated from \Cref{sec_veh_meas}, we attempt to draw lessons of general interest for wireless navigation with \glspl{ann} in \Cref{sec_discussion}.

\subsubsection{Fully-Connected Networks}

The simplest \gls{ann} structures are \glspl{fcnn}, in which all layers are flattened into one-dimensional structures and adjacent layers have direct connections among all nodes. For the wiometrics of \Cref{sec_ch_foot}, this vectorization entails equal treatment of all input bins, and any apparent differences in the order of vectorization (drawing) are merely different drawings of the same graph with all nodes in adjacent layers having connecting edges. The manner in which the \gls{csi} rows and columns were interleaved to create the \gls{ccsi} matrix becomes irrelevant, for example. The same is true for the azimuthal-frequency bins of the \gls{mfad} representation, every angular-delay bin will be represented.

Previous work on the subject has shown markedly worse performance for \glspl{fcnn} compared to similarly-sized networks employing parameter sharing \cite{chin2020intelligent}. In this work, this is tested again employing \glspl{fcnn} that are given a moderate advantage to parameter-sharing architectures in terms of the total number of weights employed. Approximately 8 million parameters are used for each of the \glspl{fcnn}. Precise network dimensions are listed in the appendix under the heading \textbf{Fully-Connected Networks}.

\subsubsection{Convolutional Neural Network}\label{cnnbeskriv}

\glspl{cnn} are learning structures that apply parallel kernels to adjacent input values through convolution operations for each convolutional layer, typically before a vectorization (flattening) operation is used for the final few fully-connected layers including the output layer. \glspl{cnn} in the \gls{mfad} domain were first suggested by Vieira et al. in \cite{joaovieira2017deep}, under the hypothesis that they should prove superior for the same reasons that \glspl{cnn} have become ubiquitous in computer vision. \glspl{cnn} that have input sizes comparable to well-known image recognition networks are used in this paper as a baseline for performance analysis. It is common practice to sub-sample ImageNet images \cite{russakovsky2015imagenet} down to 256x256 or similar, for example, and the parameters in \Cref{fig:chrepfig} are tuned accordingly. Specific network sizes are listed in the appendix, but the models employed in this work are comparatively modest in terms of trainable parameters. Approximately 4 million parameters are used as a baseline for establishing \gls{cnn} performance, with an additional round of experiments testing around 40 million parameters. They are listed under the headings \textbf{Convolutional Networks (Small)} and \textbf{Convolutional Networks (Large)} respectively in the appendix. This is far fewer parameters than the hundreds of millions employed by popular ImageNet winners for classification to tens of thousands of categories.

\section{Vehicle Measurements}\label{sec_veh_meas}
\subsection{LTE  Signals}

\begin{figure}[ht]
	\centering
	\includegraphics[width=8.5cm]{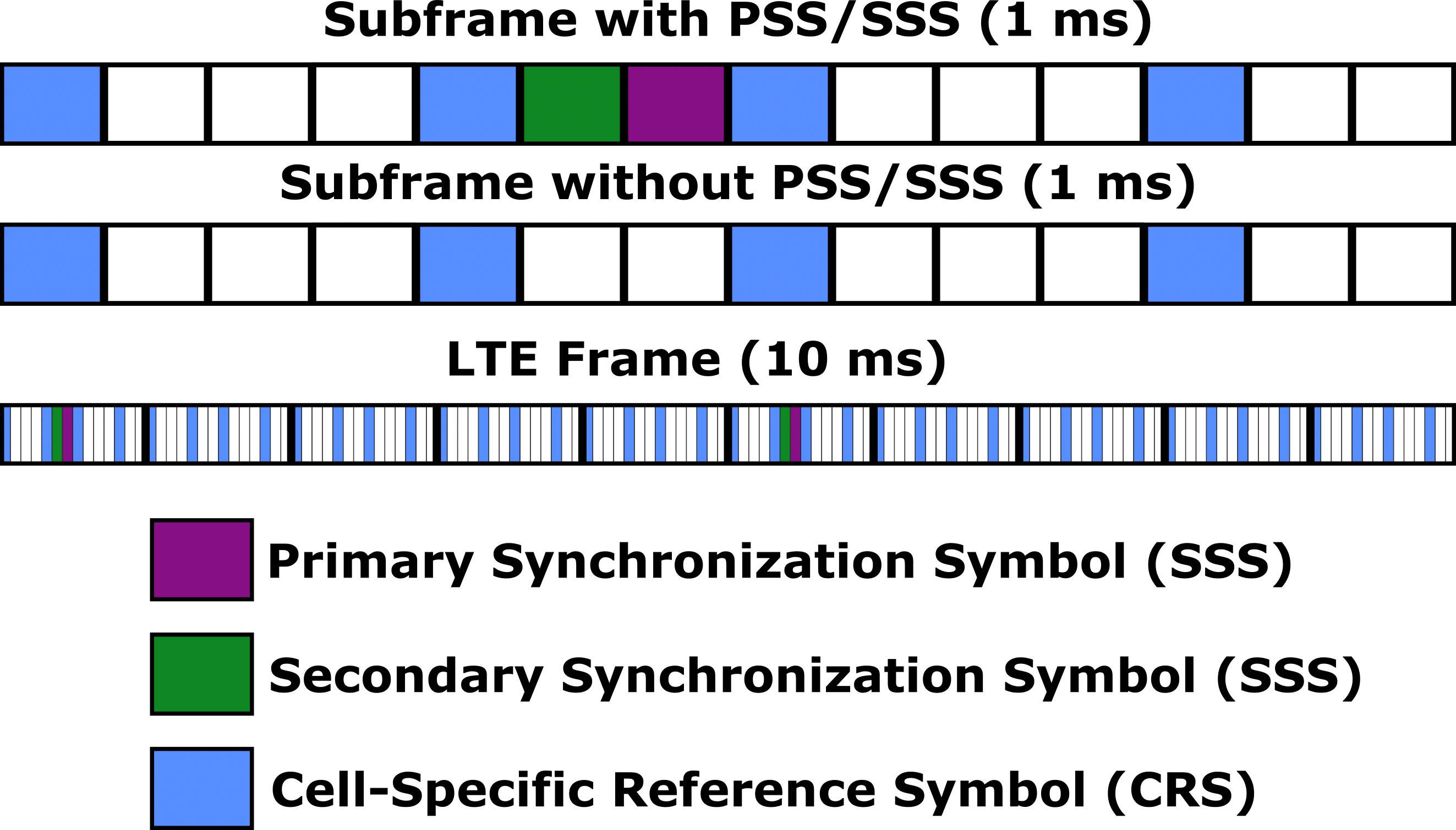}
	\caption{LTE Symbol timing for downlink reference symbols. CRS are broadcast most frequently and, though the frequency dimension of the LTE resource grid is not illustrated, span the whole channel bandwidth, offering better time resolution and more frequent observation than PSS/SSS.}
	\label{fig:lte_symbol_timing}
\end{figure}

The signals in LTE are split into frames and subframes. A number of synchronization symbols are defined that are necessary for a \gls{ue} to acquire basic network information and to perform coherent data demodulation \cite{dahlman20134g}. The idea of using these symbols for multilateration (a pseudorange model similar to GNSS) has been explored by multiple research groups \cite{muller2016statistical,driusso2016vehicular, wang2020performance}. For a detailed description of how such symbols can be acquired and used, readers are referred to \cite{shamaei_theory_to_practice}. The signals of interest, which can be decoded opportunistically (without network collaboration or knowledge) are the \gls{pss}, \gls{sss}, and \gls{crs}. \Cref{fig:lte_symbol_timing} shows how these symbols are structured in \gls{lte} \gls{fdd} subframes and frames.

\gls{crs} are transmitted with the highest frequency and span a larger bandwidth than \gls{pss}/\gls{sss}, so they offer better time resolution and more frequent observations for channel estimation. Explicit signals for positioning were introduced in \gls{lte} release 9, but are not widely deployed by network operators. 


\subsection{Measurement System}

\begin{figure}[ht]
	\centering
	\includegraphics[width=8.5cm]{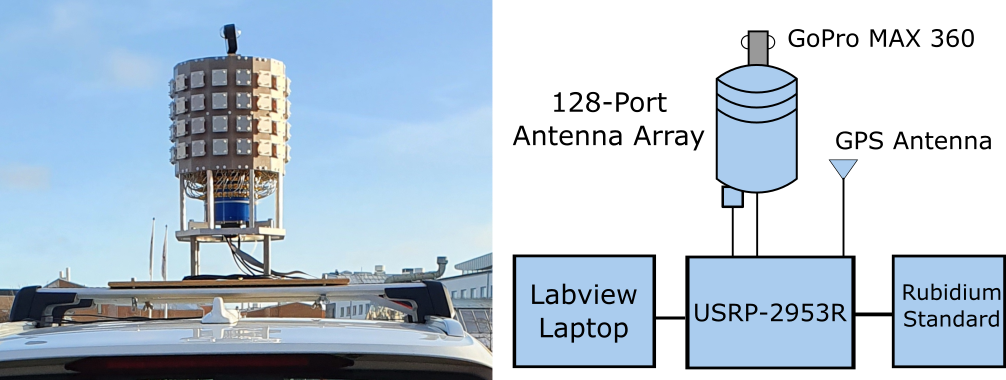}
	\caption{Picture and block diagram of LTE measurement system. A GoPro MAX panoramic camera was used to capture video of the environment as seen by the antenna array, and an OXTS 3003G system was used for Ground Truth pose estimates.  All systems held time synchronization with GPS Time as a time base.}
	\label{fig:meas_system_blockdiagram}
\end{figure}


The primary components for our measurement system are shown together with a photograph in \Cref{fig:meas_system_blockdiagram}. The backbone of the measurement system is a Software-Defined Radio (SDR) from National Instruments, USRP-2953R, together with the LabVIEW Communications Application Framework. A \gls{suca} antenna with 128 ports (64 dual-polarized patch antennas in four layers of 16 elements each) was mounted on top of the vehicle. Such arrays are used in channel sounding \cite{rimax_richter} and provide significant array gain and angular resolution, particularly in the azimuthal domain. After network acquisition of Base Station A, samples in the time domain of PSS, \gls{sss}, and \gls{crs} were continuously logged on a laptop for post-processing and subsequent channel estimation. A rubidium frequency standard, disciplined by GPS prior to the measurements, was used to provide a highly stable time reference . It should be noted that the measurement system switches at the speed of two \gls{crs} transmissions per antenna (see \Cref{fig:lte_symbol_timing}); with four symbols per subframe, that means at least 64 subframes are required to cycle among all antennas. Sampling on all antennas (one snapshot) is completed every 75 ms, allowing time for adjusting the gain control in the receiver. This sequential sampling of antennas implements sharp practical limitations on driving speed, because even with a speed of 1.0 m/s, the array moves half a wavelength over one snapshot. This limitation is applicable to our single RF-receiver measurement set-up, but could be negated with a more complicated measurement system employing parallel RF receive chains.

Ground truth estimates of vehicle position and orientation are generated using a dual-antenna, dual-constellation high-precision GNSS-Inertial system from OXTS, the RT3003G, which advertises 1 cm position accuracy and 0.05$^\circ$ heading accuracy, though these data sheet values might not be fully realized with long and slow trajectories through obstructed-sky environments. Post-Processed RTK was performed using observation data from the SWEPOS network  to give the best possible pose estimates. Panoramic video from a GoPro MAX camera mounted on the antenna array was used for manual inspection in the event of unexpected results from the other systems.

\subsection{Test Route}

\begin{figure}[ht]
	\centering
	\includegraphics[width=8.5cm]{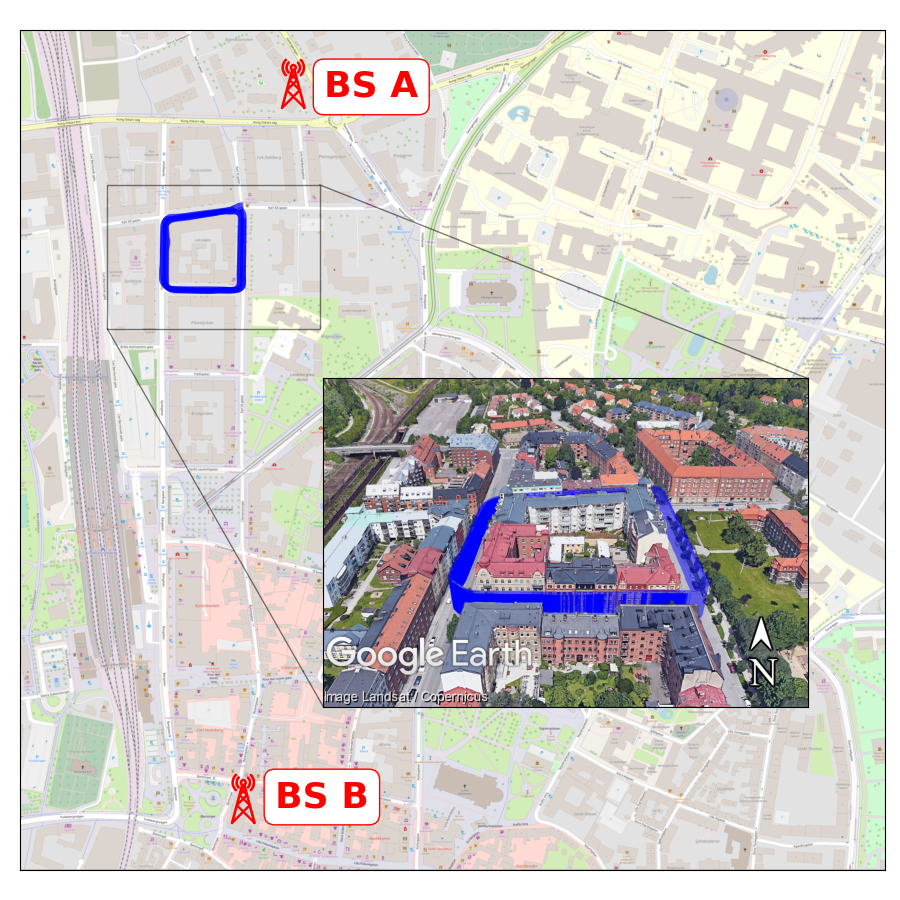}
	\caption{The two commercial base stations in Lund, Sweden (55.71$^\circ$N, 13.19$^\circ$E) and the drive route for the four laps. The inset plot shows the urban canyon three-dimensional geometry of the route, courtesy of Google Earth. Note that base station locations are not used for navigation.}
	\label{fig:bs_zoomout}
\end{figure}

A test route was driven in an urban canyon environment in downtown Lund, Sweden. Four laps were driven (starting point 55.71055$^\circ$N, 13.18919$^\circ$E). Two laps were driven in the counterclockwise direction and two were driven clockwise. The commercial base stations and their locations relative to the test vehicle driving route are shown in \Cref{fig:bs_zoomout}, together with the inset plot showing the three-dimensional building geometry. The total distance traversed for each lap was about 400 m, spanning just over 100 m East-West and North-South. The average driving speed\footnote{As stated in the previous subsection, the switching of the array is done at the rate of two \gls{crs} per antenna, which governs the channel coherence time and subsequent limitation on speed. A parallel receive apparatus would not have such a limitation but represents a significant increase in measurement complexity.} was 1.0 m/s with a total of about 22,000 snapshots, each with a duration of 75 ms. 

The commercial LTE \glspl{bs} are from a single operator broadcasting at a center frequency of 2.66 GHz ($\lambda = 11 \ cm$). A single sector (cell ID 376) of the the primary transmitter BS A is used as the only data input for all of the single base station (-S-) networks listed in the appendix. One sector (179) of a second base station is used as the additional input for the double base station (-D-) networks\footnote{In a parametric navigation problem, this would entail improved \gls{gdop}, but the concept is not directly translatable here. Intuitively, one would expect more information to result in better performance.}. BS A is not physically visible for any section of the route, it is entirely non-\gls{los}. BS B is physically visible in the Eastern section of the route at a distance of about 650-750 m. It provides at least one strong \gls{los} path for most of the Eastern section and limited non-\gls{los} energy for some other sections of the route.  

The Eastern and Western sections of the drive route are multi-lane roads with offsets in the lateral direction (lane-center to lane-center) of at least 3 m for opposing driving directions. The Northern and Southern sections are only a single car width with parked vehicles on each side.

\begin{figure}[ht]
	\centering
	\includegraphics[width=8.5cm]{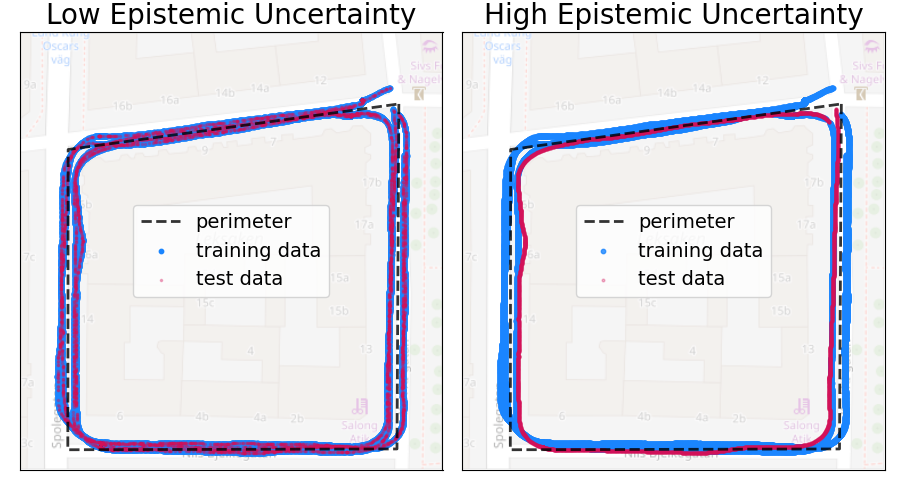}
	\caption{Two test/train splits representing very different levels of epistemic uncertainty with approximately the same number of training and test points in each. In the ``Low Epistemic Uncertainty" (left) plot, a randomly-chosen 25\% of data points from the four laps are reserved for testing. In the ``High Epistemic Uncertainty" (right) plot, one of the four laps is reserved for testing. The approximate perimeter of the route is also shown.}
	\label{fig:epistemic_craziness}
\end{figure}

The data are split into training sets in two different ways as shown in \Cref{fig:epistemic_craziness} together with the approximate perimeter of the route. First, in the \gls{leu} test case, the complete data from all four laps are pooled together and a random subset of 25\% of the whole measurement series is reserved for testing. Next, in the \gls{heu} test case, one of the four laps is reserved as a test set while the other three laps are used as training data. Having a dedicated test lap (the \gls{heu} case) entails having combinations of position and heading in testing that deviate from all training data by several meters and tens of degrees. Additionally, starts and stops occur at different locations and the scatterering environment varies with traffic in a way not fully captured in any training data; other cars and buses appear at different sections of the drive route with each loop. Additionally, only one of the three \gls{heu} training laps represents the vehicle traveling in the same direction as the test data, or in the same lane for the two-lane sections. The \gls{leu} training set includes drive data in both directions, and it is unlikely that relevant objects in the dynamic traffic environment (a passing bus, for example) will not be captured in any training samples.

\begin{remark}\label{probe_sourced_mapping_remark}
There are a few general challenges for feature-matching systems employing \glspl{ann}, primarily the necessity for large quantities of annotated training data \cite{fingerprint_tutorial_2020}. Additionally, such systems can entail significant computational burden at run-time owing to the number of operations necessary for each inference. Vehicular systems (the vehicular test platform for this work is described in \Cref{sec_veh_meas}) are well-suited to overcome these challenges, as they are designed with operation at scale in mind. Data collection can be distributed among an entire fleet of vehicles to achieve wide-scale coverage for building high-definition map databases \cite{liu2020high, HD_Map_Update_Rate}. The rest of the sensor suite provides opportunity for generating labels even in the absence of GNSS availability, allowing for the natural integration of a terrestrial wireless navigation sensor in the suite \cite{whiton2022cellular}. Furthermore, deep-learning-based perception and scene understanding from cameras is the norm in such applications \cite{guo2021survey}, meaning that real-time processing power requirements are already dimensioned for advanced \glspl{ann}.
\end{remark}



\section{Results}\label{sec_results}
\subsection{Channel Results}

\begin{figure}[ht]
	\centering
	\includegraphics[width=8.5cm]{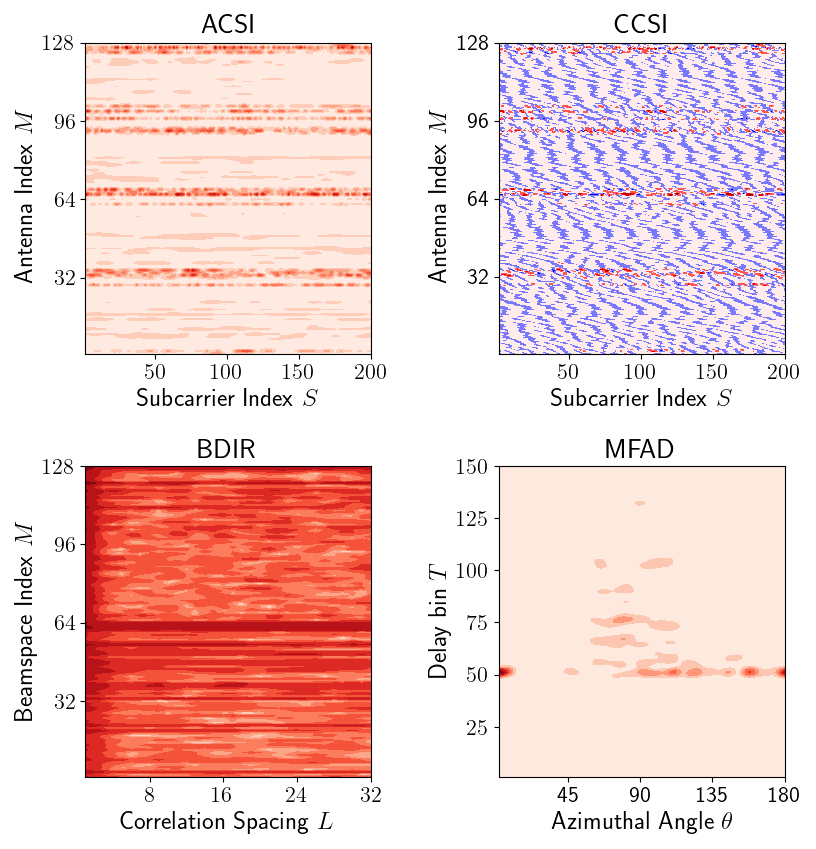}
	\caption{Visualization of the same channel measurement in the four wiometrics for snapshot index 11,000, in the Southeast corner of the route. Note that CCSI uses a color scheme that permits for negative values, (shown as increasingly blue for higher negative values).}
	\label{fig:ch_plots}
\end{figure}

A single channel measurement (derived from CSI values, see \Cref{fig:chrepfig}) is shown as each of the four wiometrics, illustrated in \Cref{fig:ch_plots}. The \gls{acsi} representation is reasonably intuitive; certain antenna ports receive significantly more energy than others, because the \gls{suca} antenna consists of elements with high directionality. There is some frequency dependence, but over the limited bandwidth of the channel it does not lend itself to intuition. The \gls{ccsi} representation shows a similar pattern but with more information than is contained by \gls{acsi}; taking absolute values of the complex CSI entails information loss. As the only representation that takes on both positive and negative values, the color scheme for \gls{ccsi} represents larger negative values as darker blue and larger positive values as darker red. The beam-domain transformation of \gls{bdir} results in clear patterns across the beam dimension, but not at the same indexes as those corresponding directly with antenna ports, as \gls{acsi} and \gls{ccsi} have. The correlation across frequencies uses only a subset of frequency offsets, representing a significant compression in input dimension compared with using all $S$ subcarriers (see the appendix for dimensions of each representation). Finally, the \gls{mfad} wiometric provides a highly intuitive spatial interpretation. Energy is received primarily from the rear and left (90 to 180 degrees in azimuth) for the first arriving path around time index 50, and then subsequent incoming energy arrives from the left side up until around time index 130.

\subsection{Baseline Performance}

\begin{figure}[ht]
	\centering
	\includegraphics[width=8.5cm]{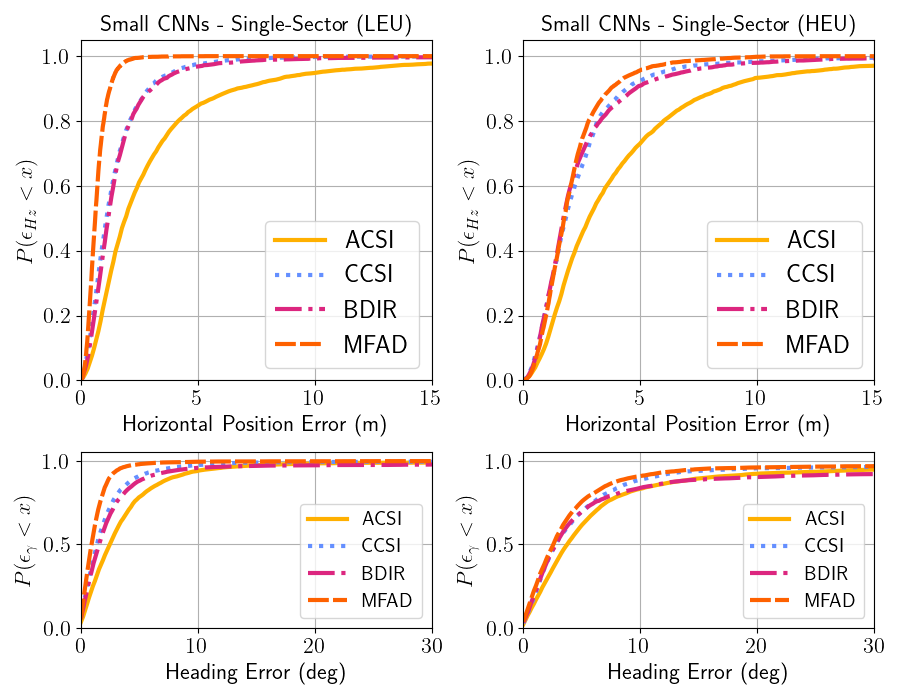}
	\caption{Position errors and heading errors for similarly-sized networks for the various channel representations with low epistemic uncertainty (left) and high epistemic uncertainty (right). See \Cref{fig:epistemic_craziness} for an illustration of the test/train split.}
	\label{fig:lil_cnn_combo}
\end{figure}

Performance is visualized through cumulative distribution functions showing horizontal position error in meters, as well as heading error spanning from $[0^\circ,180^\circ]$. Selected values for all combinations are also included in Tables \ref{tab:pos_err_tab} and \ref{tab:heading_err_tab}. A baseline comparison of the wiometrics using \glspl{cnn} (see the appendix under the heading \textbf{Convolutional Networks (Small)}) is shown in \Cref{fig:lil_cnn_combo}. Position errors and heading errors are both higher for the \gls{heu} case where the test set includes combinations of position and heading that are less representative of the training set. The \gls{mfad} representation performs best for both position and heading for the \gls{leu} and \gls{heu} cases, but performs only marginally better than the other wiometrics for the \gls{heu} case. \gls{acsi} performs worse than \gls{ccsi}, indicating that the information loss incurred when discarding phase information negatively impacts performance. The \glspl{cnn} seem to be capable of exploiting phase information and making inferences about position and heading.

\begin{figure}[ht]
	\centering
	\includegraphics[width=8cm]{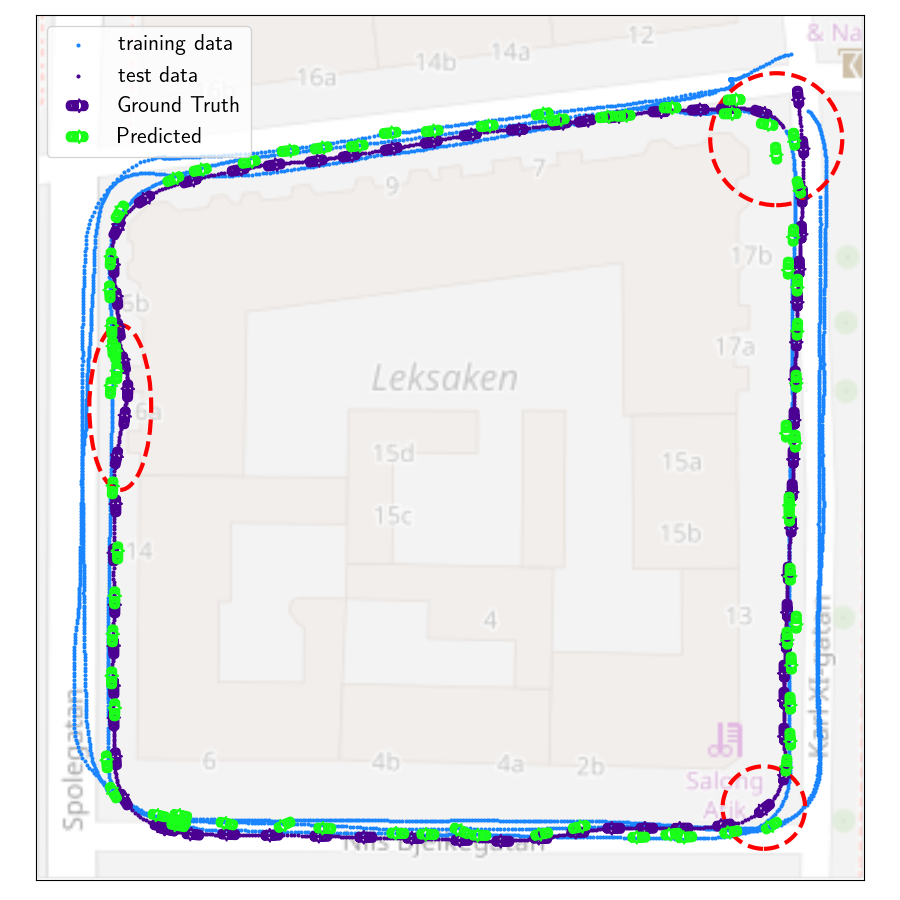}
	\caption{Predictions and ground truth for the MFAD representation CNN-MFAD-S-3.8M \gls{heu} overlaid on a map for the high epistemic uncertainty case. Note that the network does not tend to project to places outside the training set (emphasized with dashed red ellipses).}
	\label{fig:high_epistemic_map}
\end{figure}

To provide additional intuition beyond the error distributions, the \gls{heu} result from \Cref{fig:lil_cnn_combo} for the \gls{mfad} representation (CNN-MFAD-S-3.8M in the appendix) is overlaid on a map in \Cref{fig:high_epistemic_map} with the training data, test data, estimated states, and their true values for the test set. Predictions are effectively mapped to the input space of the training set, but the network does not extrapolate beyond the exterior of the training data bounds. Three sections where this is particularly visible are emphasized with red ellipses. In the Southeast corner and Western section the vehicle cuts farther to the inside than was experienced in the training set, and predictions are mapped closer to points from training. In the Northeast corner, the first few points of the test set are not covered by the training set, but are projected to the wrong side of the training data (inside rather than outside).  




Repeating the experiment using different training laps yielded similar results in terms of error distributions as well as the underlying pattern of effectively projecting test data to data in the training set. While this might be an indication of over-fitting, using a smaller number of training epochs or smaller learning rate did not decrease the error or improve the ability of the network to extrapolate effectively toward the correct position and heading. 



\subsection{Network Size and Architecture}

\begin{figure}[ht]
	\centering
	\includegraphics[width=8.5cm]{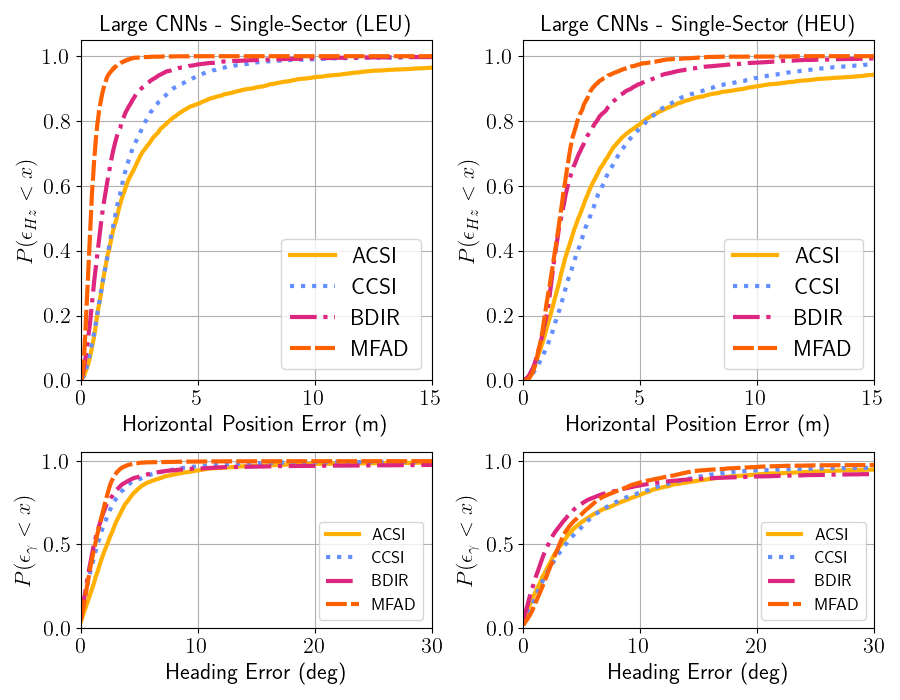}
	\caption{Position errors and heading errors for the four representations with larger convolutional neural networks. Each network has tens of millions of parameters, approximately an order of magnitude more than is shown in \Cref{fig:lil_cnn_combo}.}
	\label{fig:ginorm_network}
\end{figure}

The experiment was repeated using approximately an order of magnitude higher number of weights in each of the networks (see the appendix under the heading \textbf{Convolutional Networks (Large)}) and the results are shown in \Cref{fig:ginorm_network}. Performance is marginally better than with the smaller networks, for the \gls{mfad} and \gls{bdir} representations, but worse especially in the error tails of the \gls{acsi} and \gls{ccsi} representations.

\begin{figure}[ht]
	\centering
	\includegraphics[width=8.5cm]{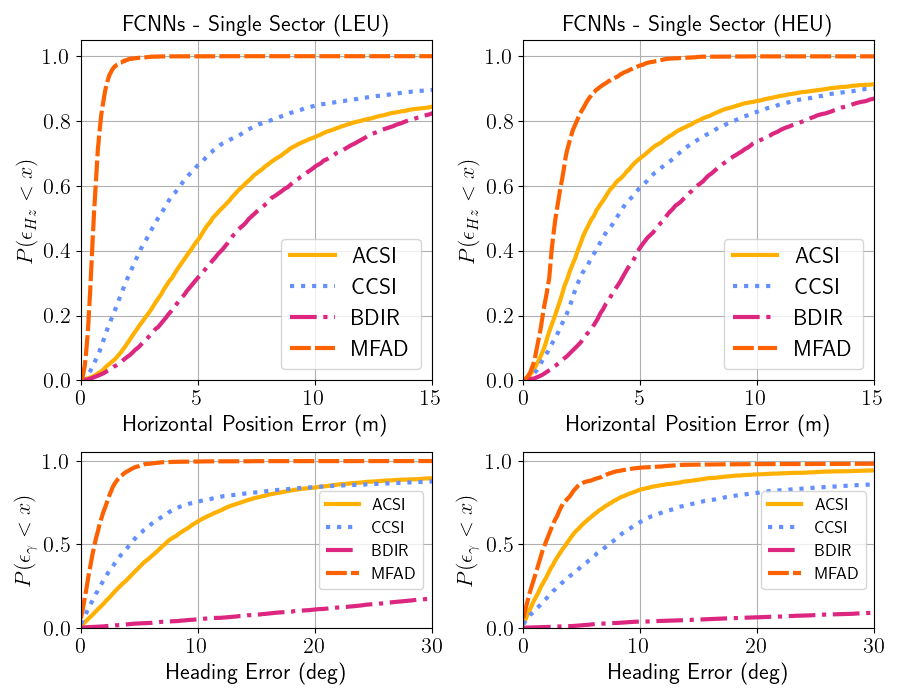}
	\caption{Position errors and heading errors for the four representations with fully-connected neural networks. Each value of the input space is treated as a particular bin which is weighted in the first layer irrespective of adjacent bins in the original form.}
	\label{fig:fcn_curves}
\end{figure}

The results of the same input data with \glspl{fcnn} (see the appendix under the heading \textbf{Fully-Connected Networks}) are shown in \Cref{fig:fcn_curves}. Performance is markedly degraded compared with the baseline \glspl{cnn} of the previous section, despite the networks having twice the number of trainable parameters, with the exception of the \gls{mfad} representation, which has similar performance. This indicates that the convolution operations are extracting useful information for all the wiometrics, but that the feature engineering going into creation of \gls{mfad} may be performing a function similar to what is being learned by the network. Heading estimates in particular are significantly degraded, and for the \gls{bdir} representation even more than for the other representations which use either antenna indexes or the complex radiation pattern. The \gls{bdir} heading performs similarly to blind sampling of a uniform distribution from $[-180^\circ,180^\circ)$. Another curiosity is that for the \gls{leu} case, \gls{acsi} performs better than \gls{ccsi}, indicating that the information loss from discarding phase information is actually beneficial, possibly because values learned for phase over the spacing of a few wavelengths tend to only contribute to overfitting, whereas the convolutional operations in the \glspl{cnn} learn more meaningful features about adjacent frequencies and antennas.

\subsection{Multiple Base Stations}\label{multiple_bs}

\begin{figure}[ht]
	\centering
	\includegraphics[width=8.5cm]{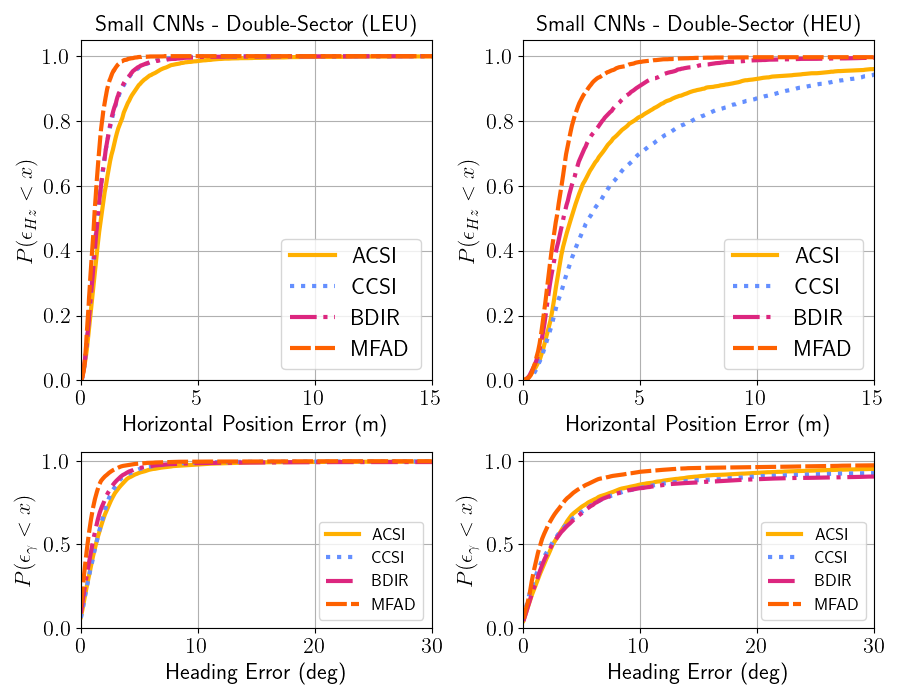}
	\caption{Position errors and heading errors for the four channel representations trained with data from two sectors and the two levels of epistemic uncertainty, one from each of BS A and BS B (see \Cref{fig:bs_zoomout}). The input data from each sector is stacked, and the convolutions are two-dimensional.}
	\label{fig:two_sector}
\end{figure}

\begin{figure}[ht]
	\centering
	\includegraphics[width=8.5cm]{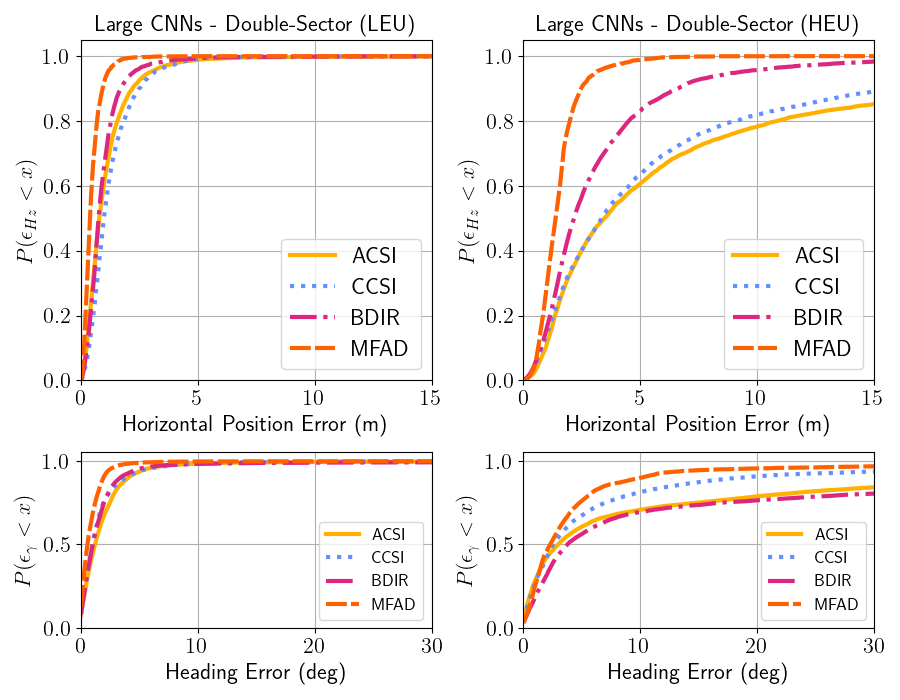}
	\caption{Position errors and heading errors for the four channel representations trained with data from two sectors and the two levels of epistemic uncertainty, one from each of BS A and BS B (see \Cref{fig:bs_zoomout}). The data from each sector is stacked, and the convolutions are two-dimensional. }
	\label{fig:two_sector_doubles}
\end{figure}

Finally, results for \glspl{cnn} utilizing both \glspl{bs} (both BS A and BS B from \Cref{fig:bs_zoomout}) are shown in \Cref{fig:two_sector} and \Cref{fig:two_sector_doubles}. For the \gls{mfad} and \gls{bdir} representations the use of the second \gls{bs} results in superior performance (except for large heading tail errors for \gls{bdir}-\gls{heu}). Addition of the second \gls{bs} leads to significantly worse performance for \gls{acsi} and \gls{ccsi} in the \gls{heu} case, even if it improves performance nominally in the \gls{leu} case. Performing three-dimensional convolutions significantly increases training time and inference at run-time, and for these representations it results in worse performance.

\subsection{Summary of Results}

Position and heading error values for the 68th, 95th, and 99th percentiles are included in \Cref{tab:pos_err_tab} and \Cref{tab:heading_err_tab}, under the network names from the appendix. Individual wiometrics are sorted into adjacent rows to facilitate comparison. In all cases it is apparent that \gls{leu} performs better than \gls{heu}; it is unsurprising that generalization to new data is difficult. Some wiometrics are clearly superior to others; specifically, \gls{mfad} outperforms the other representations in all cases, while \gls{acsi} tends to perform the worst in most cases. 

\begin{table}[ht]
    \centering
    \caption{Table of Horizontal Position error values}
    \begin{tabular}{c|c|c|c|c|c|c} 
      Network Name & \multicolumn{3}{c|}{LEU Error (m)} & \multicolumn{3}{c}{HEU Error (m)}\\ \hline
      Error Percentile (\%) & 68 & 95 & 99 & 68 & 95 & 99 \\ \hline
        FCNN-ACSI-S-7.5M & 8.2 & 58 & 92 & 4.9 & 25 & 94 \\
        CNN-ACSI-S-3.9M & 2.9 & 10 & 20 & 4.3 & 12 & 21 \\
        CNN-ACSI-S-30.6M & 2.4 & 12 & 25 & 3.4 & 16 & 33 \\
        CNN-ACSI-D-4.5M & 1.2 & 3.2 & 5.7 & 3.2 & 13 & 36 \\
        CNN-ACSI-D-33.0M & 1.1 & 2.9 & 5.1 & 6.2 & 34 & 68 \\
        \hline
        FCNN-CCSI-S-7.3M & 5.2 & 37 & 86 & 6.3 & 28 & 86 \\
        CNN-CCSI-S-3.8M & 1.6 & 3.9 & 6.7 & 2.6 & 5.9 & 12 \\
        CNN-CCSI-S-27.0M & 2.0 & 5.3 & 9.7 & 3.9 & 11 & 21 \\
        CNN-CCSI-D-4.0M & 1.0 & 2.2 & 3.4 & 4.7 & 16 & 25 \\
        CNN-CCSI-D-31.0M & 1.4 & 3.1 & 4.8 & 5.6 & 21 & 36 \\ 
        \hline
        FCNN-BDIR-S-7.5M & 10 & 30 & 74 & 8.8 & 23 & 64 \\
        CNN-BDIR-S-3.3M & 1.6 & 4.0 & 8.6 & 2.4 & 6.7 & 13 \\
        CNN-BDIR-S-39.9M & 1.3 & 3.6 & 8.2 & 2.3 & 6.4 & 13 \\
        CNN-BDIR-D-3.8M & 1.0 & 2.2 & 3.7 & 2.4 & 6.2 & 11 \\ 
        CNN-BDIR-D-35.7M & 1.0 & 2.2 & 4.3 & 3.9 & 9.8 & 17 \\
        \hline
        FCNN-MFAD-S-7.7M & 0.7 & 1.3 & 2.0 & 1.8 & 4.3 & 6.0 \\
        CNN-MFAD-S-3.8M & 0.8 & 1.5 & 2.1 & 2.3 & 4.8 & 8.0 \\
        CNN-MFAD-S-30.8M & 0.6 & 1.2 & 2.0 & 1.9 & 4.0 & 6.2 \\
        CNN-MFAD-D-3.8M & 0.7 & 1.4 & 2.0 & 1.8 & 3.5 & 5.9 \\
        CNN-MFAD-D-30.9M & 0.5 & 1.1 & 1.7 & 1.7 & 3.1 & 5.1 \\
        \hline
    \end{tabular}
    \label{tab:pos_err_tab}
\end{table}

\begin{table}[ht]
    \centering
    \caption{Table of Heading error values}
    \begin{tabular}{c|c|c|c|c|c|c}
      Network Name & \multicolumn{3}{c|}{LEU Error (deg)} & \multicolumn{3}{c}{HEU Error (deg)}\\ \hline
      Error Percentile (\%) & 68 & 95 & 99 & 68 & 95 & 99 \\ \hline
        FCNN-ACSI-S-7.5M & 11 & 60 & 134 & 5.9 & 35 & 129 \\
        CNN-ACSI-S-3.9M & 3.8 & 11 & 30 & 5.7 & 32 & 122 \\
        CNN-ACSI-S-30.6M & 3.2 & 11 & 31 & 5.9 & 31 & 124 \\
        CNN-ACSI-D-4.5M & 2.1 & 6.0 & 14 & 4.1 & 29 & 114 \\
        CNN-ACSI-D-33.0M & 2.0 & 5.4 & 14 & 7.9 & 92 & 146 \\ \hline
        FCNN-CCSI-S-7.3M & 7.2 & 92 & 170 & 11 & 110 & 174 \\
        CNN-CCSI-S-3.8M & 2.1 & 6.9 & 15 & 4.6 & 18 & 111 \\
        CNN-CCSI-S-27.0M & 2.3 & 8.0 & 17 & 6.1 & 23 & 121 \\
        CNN-CCSI-D-4.0M & 1.9 & 4.8 & 10 & 4.5 & 52 & 142 \\ 
        CNN-CCSI-D-31.0M & 1.5 & 5.5 & 12 & 5.0 & 42 & 146 \\ \hline
        FCNN-BDIR-S-7.5M & 97 & 157 & 174 & 134 & 170 & 178 \\
        CNN-BDIR-S-3.3M & 2.3 & 8.8 & 102 & 4.7 & 71 & 170 \\
        CNN-BDIR-S-39.9M & 1.8 & 8.8 & 116 & 3.8 & 60 & 163 \\
        CNN-BDIR-D-3.8M & 1.4 & 4.5 & 14 & 4.9 & 120 & 178 \\
        CNN-BDIR-D-35.7M & 1.6 & 4.9 & 23 & 23 & 148 & 176 \\ \hline
        FCNN-MFAD-S-7.7M & 1.8 & 4.2 & 6.7 & 2.9 & 9.0 & 71 \\
        CNN-MFAD-S-3.8M & 1.3 & 3.2 & 6.9 & 3.9 & 15 & 107 \\
        CNN-MFAD-S-30.8M & 1.8 & 3.3 & 5.5 & 4.9 & 17 & 98 \\
        CNN-MFAD-D-3.8M & 0.9 & 2.9 & 6.3 & 2.5 & 13 & 102 \\
        CNN-MFAD-D-30.9M & 1.0 & 2.5 & 5.3 & 3.8 & 17.6 & 94.4 \\
        \hline
        
    \end{tabular}
    \label{tab:heading_err_tab}
\end{table}





\section{Discussion}\label{sec_discussion}
The results demonstrate that a single commercial \gls{lte} \gls{bs} in non-\gls{los} conditions can be used to attain meter-level horizontal positioning accuracy as well as heading accuracy ranging from a few degrees to tens of degrees by employing deep learning architectures that are well-established in computer vision. This is achieved by using the richness of the multipath channel in a complicated urban propagation environment populated with irregular surfaces and variation in the geometry of scatterers. Even with limited training data and varied channel representations (wiometrics) ranging from bulky ones with minimal processing (\gls{acsi} and \gls{ccsi}) to more intricate feature engineering (\gls{bdir} and \gls{mfad}), both position and heading estimates perform well on a snapshot-by-snapshot basis without prior knowledge of navigation states.  

As originally hypothesized by Vieira et al. in \cite{joaovieira2017deep}, the feature engineering involved in generating the \gls{mfad} representation offers compelling performance. This representation takes full advantage of consistent clock synchronization between sender and receiver and integrates the complete radiation pattern of the receive antenna array. The representation is so effective that performance was essentially the same even when employing a simpler \gls{fcnn} network architecture, when adding the information provided by a second \gls{bs}, or using a convolutional network with an order of magnitude more training weights. None of the other wiometrics achieve the same performance under any circumstances, and, consistent with \cite{chin2020intelligent}, \glspl{fcnn} are inferior to \glspl{cnn} for all cases. Particularly for \gls{bdir}, the convolution operations appear to be critical to attaining reasonable performance. The network architecture is more consequential for the simpler representations, and it is worthwhile to explore more intricate network architectures (particularly Transformers) for these, or to explore the use of Transformers for the integration of multiple \glspl{bs}.

The difference in performance between the \gls{heu} and \gls{leu} training and testing splits is pronounced for all channel representations and network architectures. The input space in this paper draws from only four passes over the same route, two in each direction, and it is reasonable to suspect that a more thorough surveying of the test area with more combinations of position and heading would result in improved performance, particularly for the error tails, which in the \gls{heu} case correspond strongly with test points that are not representative of the training set.

There is significant room for wiometric navigation literature to explore robustness to environmental changes in a manner similar to what has been done in computer vision for positioning \cite{sattler2018benchmarking}. While a larger training set as compared to the \gls{heu} case visualized in \Cref{fig:epistemic_craziness} would presumably improve performance, multiple parameters are still not challenged. All testing was done in the winter and vegetation will change, creating a different scattering environment seasonally. Some non-parametric wireless positioning literature considers measurements taken spaced over the course of months \cite{ferrand2020dnn}. In our data collection, there were few changes in the arrangement of cars parked along the side of the road and traffic density did not change significantly even if certain scatterers (buses and other vehicles) were variable from lap to lap. More critically for the \gls{ccsi} and \gls{mfad} representations, absolute time synchronization is effectively maintained between transmitter and receiver, which requires first an accurate initialization and then an oscillator on both sides with an Allan deviation not representative of technologies used in today's commercial devices (see \cite{teunissen2017springer} Fig. 5.4).

\section{Conclusion}\label{sec_conclusion}
In this manuscript, a massive-MIMO antenna and receiver system was mounted on a passenger vehicle to opportunistically receive downlink \gls{crs} from commercial LTE \glspl{bs} and to use them to generate two-dimensional estimates of position and heading from wiometrics using various deep learning architectures. In non-\gls{los} conditions without knowledge of the transmitter location, meter-level estimates of position and heading errors of a few degrees are realized.

Given the superiority of the \gls{mfad} representation, future work should explore some of the practical challenges involved in realizing this at scale, including maintaining stable clock synchronization, designing antenna arrays with high spatial resolution that can be integrated into the form factor of a vehicle, and addressing calibration for the receive chains.

Another line of future work is to improve performance by exploring alternative network architectures (Transformers in particular) and testing the environmental conditions under which performance can be shown to be variable in order to attain more robust navigation estimates. Alternative wiometrics beyond those explored in this manuscript may also be useful, and performance in different environments and with different channel bandwidths, carrier frequencies, and measurement apertures are also interesting. Considering that the results shown in \Cref{sec_results} are achieved with \gls{lte} signals that have limited bandwidth, it seems likely that statistical methods will out-perform state-of-the-art parametric methods for many problems in wireless navigation as they have in computer vision.



\section*{Acknowledgment}
The authors would like to thank Martin Nilsson at Lund University for his help in setting up the measurement system.
\section*{Appendix A}

Network names follow the following convention XXX-YYYY-B-NNM. Where: 
\begin{itemize}
    \item XXX is either \gls{cnn} (convolutional) or \gls{fcnn} (fully-connected)
    \item YYY is one of the four-letter wiometrics (see \Cref{fig:chrepfig})
    \item B is either S for single base station or D for double base station
    \item NN is the approximate number of weights of the network, rounded to the nearest 100,000 and expressed in millions (M)
\end{itemize}

Each network shape is randomly initialized and trained twice, once each for the \gls{leu} or \gls{heu} training and test sets. 

\begin{table*}[ht]
    \centering
    \caption{Table of Network values}
    \begin{tabular}{c|c|c|c|c}
      Network Name & Input Shape & Layers & Convolutional Filters & Fully-Connected Layer Nodes \\ \hline
      \multicolumn{5}{l}{\textbf{Fully-Connected Networks}}\\
      \hline
      FCNN-ACSI-S-7.6M & 200 $\times$ 128 & 7 &   & 256, 1024, 512, 256, 256, 128, 64 \\
      FCNN-CCSI-S-7.6M & 200 $\times$ 256 & 7 &   & 128, 256, 1024, 512, 256, 256, 128 \\
      FCNN-BDIR-S-7.6M & 32 $\times$ 128 & 7 &   & 1024, 2048, 512, 256, 256, 128, 64 \\
      FCNN-MFAD-S-7.4M & 150 $\times$ 90 & 7 &   & 512, 512, 256, 256, 128, 128, 64 \\
      \hline
      \multicolumn{5}{l}{\textbf{Convolutional Networks (Small)}}\\
      \hline
      CNN-ACSI-S-3.9M & 200 $\times$ 128 & 8 &  16, 32, 64, 128 & 256, 1024, 256, 128 \\
      CNN-CCSI-S-3.8M & 200 $\times$ 256 & 8 &  16, 32, 64, 128 & 128, 1024, 256, 128 \\
      CNN-BDIR-S-3.3M & 32 $\times$ 128 & 7 &  16, 32, 64 & 512, 1024, 512, 128 \\
      CNN-MFAD-S-3.8M & 150 $\times$ 90 & 7 &  16, 32, 64 & 256, 1024, 256, 128 \\
      \hline
      \multicolumn{5}{l}{\textbf{Convolutional Networks (Large)}}\\
      \hline
      CNN-ACSI-S-30.6M & 200 $\times$ 128 & 8 &  32, 64, 128, 256 & 1024,2056,1024,512 \\
      CNN-CCSI-S-27.0M & 200 $\times$ 256 & 8 &  32, 64, 128, 256 & 512, 1024, 512, 256 \\
      CNN-BDIR-S-39.9M & 32 $\times$ 128 & 7 &  32, 64, 128 & 2048, 5096, 2048, 1024 \\
      CNN-MFAD-S-30.8M & 150 $\times$ 90 & 7 &  32, 64, 128 & 1024, 2048, 1024, 512 \\
      \hline
      \multicolumn{5}{l}{\textbf{Double Base Station Networks (Small)}}\\
      \hline
      CNN-ACSI-D-4.5M & 200 $\times$ 128 $\times$ 2 & 8 &  16, 32, 64, 128 & 1024, 1024, 256, 128 \\
      CNN-CCSI-D-4.0M & 200 $\times$ 256 $\times$ 2 & 8 &  16, 32, 64, 128 & 512, 1024, 256, 128 \\
      CNN-BDIR-D-3.8M & 32 $\times$ 128 $\times$ 2  & 7 &  16, 32, 64 & 1024, 1024, 512, 128 \\
      CNN-MFAD-D-3.8M & 150 $\times$ 90 $\times$ 2  & 7 &  16, 32, 64 & 512, 1024, 256, 128 \\
      \hline
      \multicolumn{5}{l}{\textbf{Double Base Station Networks (Large)}}\\
      \hline
      CNN-ACSI-D-33.0M & 200 $\times$ 128 $\times$ 2 & 7 &  32, 64, 128 & 2048, 2048, 1024, 512 \\
      CNN-CCSI-D-31.0M & 200 $\times$ 256 $\times$ 2 & 7 &  32, 64, 128 & 1024, 2048, 1024, 512 \\
      CNN-BDIR-D-35.7M & 32 $\times$ 128 $\times$ 2  & 7 &  32, 64, 128 & 4096, 4096, 2048, 1024 \\
      CNN-MFAD-D-31.9M & 150 $\times$ 90 $\times$ 2  & 7 &  32, 64, 128 & 2048, 4096, 2048, 1024 \\
    \end{tabular}
    \label{tab:Requirements_n_Benefits}
\end{table*}

A kernel size of 4 is used for convolutions. Max pooling is applied after convolution and activation with a pool size of 2 $\times$ 2 for single base station two-dimensional convolutions and 1 $\times$ 2 $\times$ 2 for double base station three-dimensional convolutions. An Adam optimizer is used for all training, and a learning rate of 0.001. Dimensioning networks for variable input shapes is not an exact science, and there are small variations in the same category to target a similar number of weights for comparison.

\ifCLASSOPTIONcaptionsoff
  \newpage
\fi

\bibliographystyle{IEEEtran}

\newpage
\bibliography{liberry.bib}
%



%

\begin{IEEEbiography}[{\includegraphics[width=1in,height=1.25in,clip,keepaspectratio]{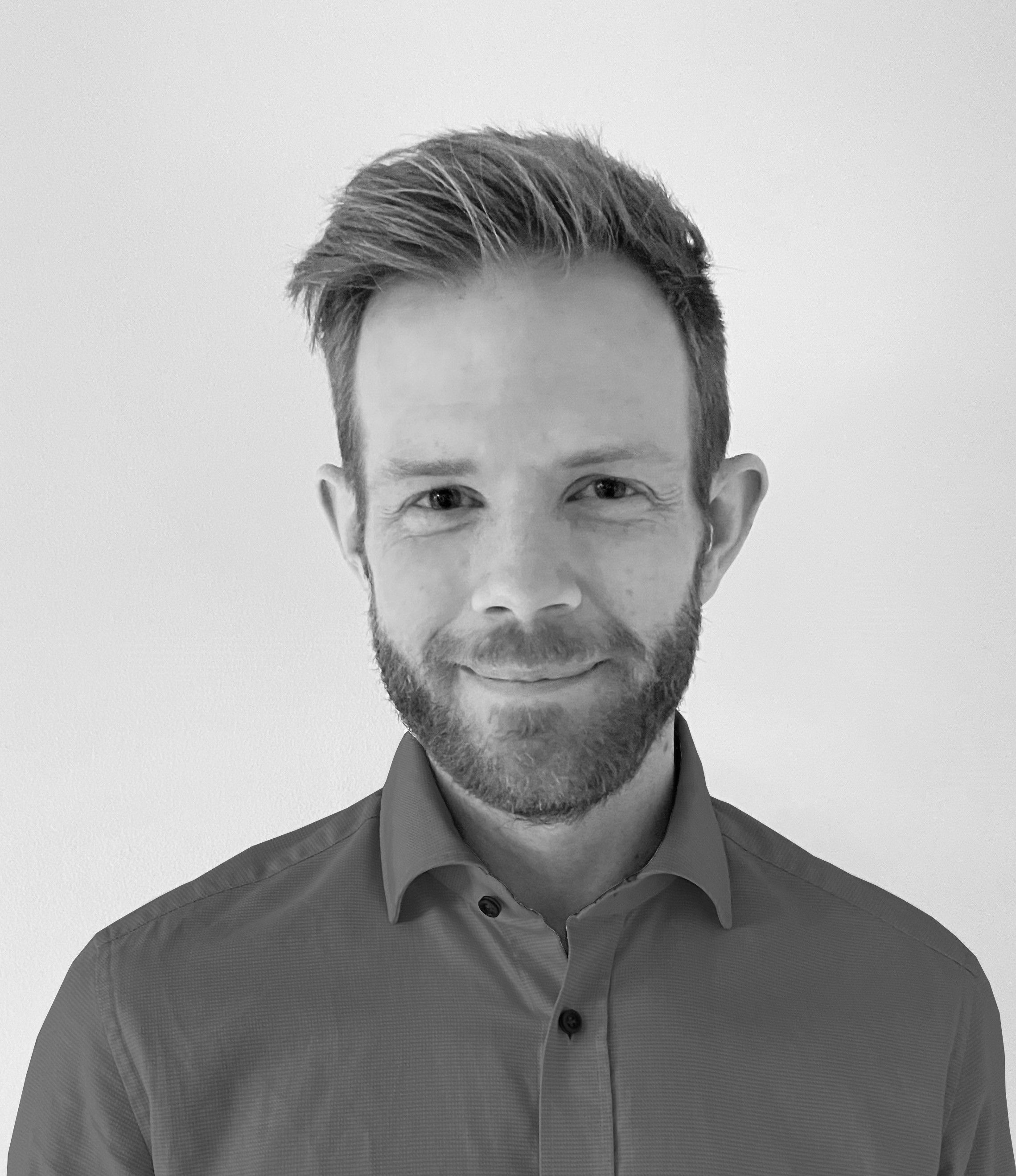}}]{Russ Whiton}~is a Student Member of IEEE.  He received an M.Sc. in electrical engineering degree from Lund University in Lund, Sweden in 2012.

He is currently employed as a Senior Hardware Engineer with Volvo Car Corporation in Gothenburg, Sweden and is a Ph.D. student at Lund University in Lund, Sweden. He has previously worked as an engineer at Volvo Group Trucks Technology, Sony Mobile Communications, and Qualcomm. His research interests include localization and wireless systems. 

Mr. Whiton is a member of the Institute of Navigation.
\end{IEEEbiography}

\vskip 0pt plus -1fil

\begin{IEEEbiography}[{\includegraphics[width=1in,height=1.25in,clip,keepaspectratio]{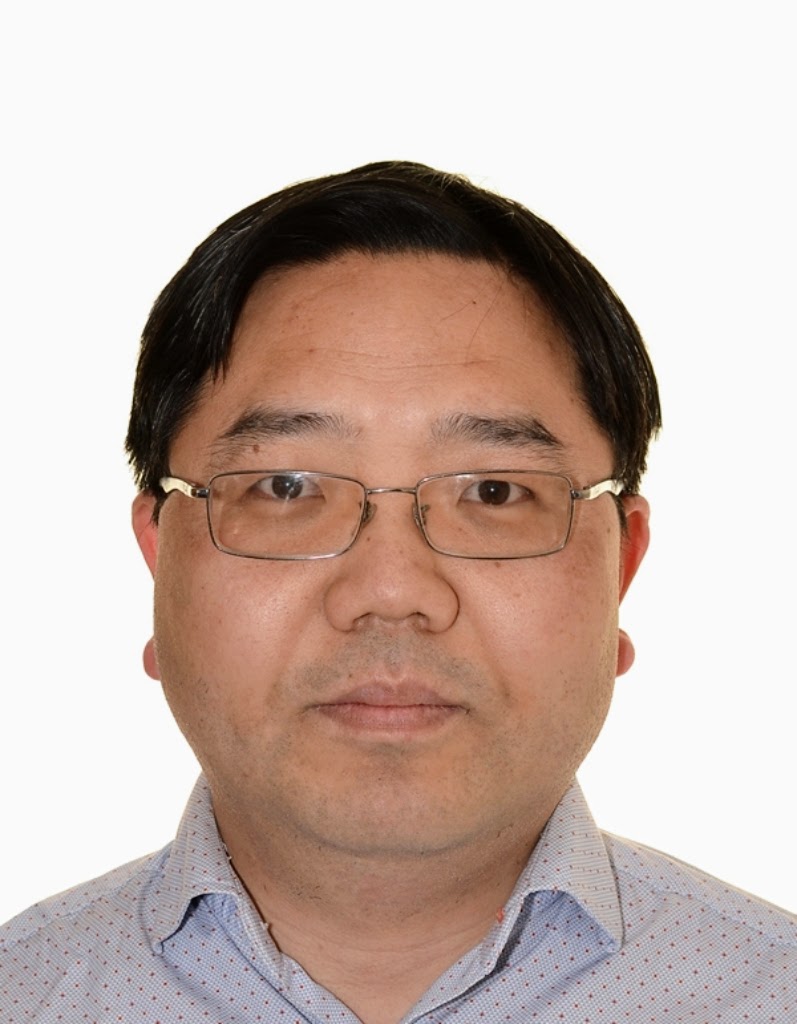}}]{Junshi Chen}~is a Student Member of IEEE. He received an M.Sc. in electrical engineering degree from Beijing Jiaotong University in Beijing, China in 2005.

He is currently employed as a Senior Algorithm Engineer with Terranet AB in Lund, Sweden and is a Ph.D. student at Lund University in Lund, Sweden. He has previously worked as a Senior Algorithm Engineer at Huawei Technologies in Sweden and China. His research interests include localization and wireless signal processing.
\end{IEEEbiography}

\vskip 0pt plus -1fil

\begin{IEEEbiography}[{\includegraphics[width=1in,height=1.25in,clip,keepaspectratio]{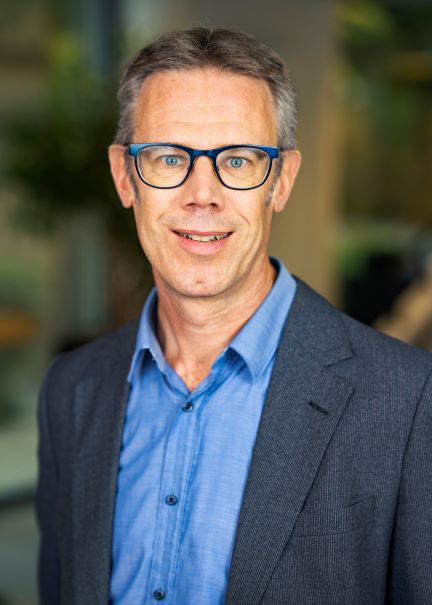}}]{Fredrik Tufvesson}~(Fellow, IEEE)~received the Ph.D. degree from Lund University, Lund, Sweden, in 2000.

After two years at a startup company, he joined the Department of Electrical and Information Technology, Lund University, where he is currently a Professor of radio systems. He has authored around 100 journal articles and 150 conference papers. His main research interest is the interplay between the radio channel and the rest of the communication system with various applications in 5G/B5G systems, such as massive multiple-input multiple-output (MIMO), mmWave communication, vehicular communication, and radio-based positioning.

Dr. Tufvesson’s research has been awarded the Neal Shepherd Memorial Award for the Best Propagation Paper in the IEEE Transactions on Vehicular Technology and the IEEE Communications Society Best Tutorial Paper Award.
\end{IEEEbiography}




\end{document}